\documentclass[11pt,hyper]{JHEP3}
\usepackage{amsmath,epsfig}
\usepackage{amssymb,amsfonts}
\usepackage{graphicx}
\usepackage{multirow,cite}
\usepackage{subfigure}
\usepackage{slashed}
\usepackage{latexsym}
\usepackage{slashed}

\newcommand{\bi}{\begin{itemize}}
\newcommand{\ei}{\end{itemize}}
\newcommand{\non}{\nonumber}
\def\p{\partial}
\def\a{\alpha}
\def\b{\beta}
\def\d{\delta}
\def\g{\gamma}
\def\l{\lambda}
\def\e{\epsilon}
\def\k{\kappa}
\def\th{\theta}
\def\om{\omega}
\def\s{\sigma}
\def\tl{\tilde{\lambda}}
\def\ve{\varepsilon}

\def\O{\mathcal{O}}

\def\G{\Gamma}

\def\L{\Lambda}

\def\tt{\tilde{t}}
\def\ty{\tilde{y}}

\def\S{\Sigma}

\def\r{\rightarrow}
\def\half{{\frac12}}

\newcommand{\N}{{\mathcal{N}}}

\newcommand{\bea}{\begin{eqnarray}}
\newcommand{\eea}{\end{eqnarray}}
\newcommand{\be}{\begin{equation}}
\newcommand{\ee}{\end{equation}}

\title{Kerr/CFT, dipole theories and nonrelativistic CFTs}
\author{Sheer El-Showk and Monica Guica
\vspace{0.5cm}

{\it Institut de Physique Th\'eorique \\ \vspace{ 0.9 mm}
 \hspace{-0.55 cm} CEA Saclay, CNRS-URA 2306,\\ 
 \hspace{-0.55 cm} 91191 Gif sur Yvette, France
} }
\abstract{

\vspace{5 mm}
We study solutions of type IIB supergravity which are $SL(2,\mathbb{R}) \times SU(2) \times U(1)^2$ 
invariant deformations of $AdS_3 \times S^3 \times K3$ and take the form of products of self-dual spacelike warped $AdS_3$ and a deformed three-sphere. One of these backgrounds has been recently argued
to be relevant for a derivation of Kerr/CFT from string theory, whereas the remaining ones are holographic duals of two-dimensional dipole theories and their S-duals. We show that each of these backgrounds is holographically dual to a deformation of the DLCQ of the D1-D5 CFT by a specific supersymmetric (1,2) operator, which we write down explicitly in terms of twist operators at the free orbifold point. The deforming operator is argued to be exactly marginal with respect to the zero-dimensional nonrelativistic conformal (or Schr\"{o}dinger) group - which is simply $SL(2,\mathbb{R})_L \times  U(1)_R$. Moreover, in the supergravity limit of large $N$ and strong coupling, no other single-trace operators are turned on. We thus propose that the field theory duals to the backgrounds of interest are nonrelativistic CFTs defined by adding the single Schr\"{o}dinger-invariant $(1,2)$ operator mentioned above to the original CFT action. Our analysis indicates that the rotating extremal black holes we study are best thought of as finite right-moving temperature (non-supersymmetric) states in the above-defined supersymmetric nonrelativistic CFT and hints towards a more general connection between Kerr/CFT and two-dimensional 
non-relativistic CFTs. }
\begin{document}

\section{Introduction}

The AdS/CFT correspondence \cite{malda,witten,gubpol} has been one of the most fruitful ideas that have emerged from string theory in recent years. The correspondence states that quantum gravity in $d+1$ dimensional anti-de Sitter space is equivalent to a conformal field theory in $d$ dimensions. The case of the $AdS_3/CFT_2$ correspondence is particularly interesting, as it provides the basis for our understanding of the microscopic entropy of many black holes in and outside of string theory \cite{Strominger:1997eq}. While the duality is expected to hold for \emph{any} consistent theory of quantum gravity, so far the best understood and most concrete examples have been in the context of string theory\footnote{Lately there has also been much progress in understanding higher spin gauge theories in AdS \cite{Vasiliev:2003ev,
Gaberdiel:2010pz} (for a more comprehensive list of references see e.g.
\cite{Papadodimas:2011pf}).
}. 

Soon after its discovery, it was realised that the correspondence could be
modified to a duality between asymptotically anti-de Sitter spacetimes and
\emph{local} quantum field theories with a UV fixed point
\cite{Akhmedov:1998vf,Boer:1999xf,Verlinde:1999xm}. Nevertheless, in recent years non-asymptotically AdS
spacetimes have started to make their appearance more and more often, especially
in attempts to connect holography with the real world. We will be interested in
two examples. 

The first example is that of the Kerr/CFT correspondence \cite{kerrcft} and its generalizations \cite{Hartman:2008pb,Lu:2008jk,Azeyanagi:2008kb,Chow:2008dp,Azeyanagi:2008dk,Isono:2008kx,Chen:2009xja,Lu:2009gj,Amsel:2009ev,Compere:2009dp,
Krishnan:2009tj,Astefanesei:2009sh,Wen:2009qc,Azeyanagi:2009wf,Peng:2009wx,matsuo}.
In its generalized form, this correspondence explains the entropy of \emph{all} extremal black holes (in four and five dimensions at least) in terms of state counting in a conformal field theory. The never-failing agreement of the Bekenstein-Hawking with the microscopic entropy is believed to be due to a universal holographic duality between the near-horizon region of extreme black holes and a conformal field theory. The part of the (four- or higher-dimensional) near-horizon geometry
most relevant for this duality seems to be the so-called \emph{self-dual spacelike warped} $AdS_3$ factor

\be
ds^2 = - r^2 dt^2 + \frac{dr^2}{r^2} + a^2 (d\varphi + r dt)^2 \;, \;\;\;\;\;\;\;\; \varphi \sim \varphi + 2 \pi \label{kerrm}
\ee
which appears universally in the near-horizon of all these black holes \cite{reallkund}. The ``warping'' parameter $a$ generally depends on the polar angle, but it has no $r$ dependence. Note that for $a=1$ this spacetime is nothing but a quotient of $AdS_3$. In general there are also overall conformal factors which depend on the polar angle. For higher-dimensional black holes the way to isolate a spacelike warped $AdS_3$ factor inside the near-horizon geometry is not unique \cite{Hartman:2008pb,Lu:2008jk,Azeyanagi:2008kb}, a fact which leads to different ``CFT'' descriptions of the black hole, which are believed to be U-dual to each other.

The main piece of supporting evidence for the Kerr/CFT conjecture is the fact that the symmetries of the phase space of gravity on the near-horizon region (defined by appropriately-chosen boundary conditions)
generate a Virasoro algebra. This so-called asymptotic symmetry group (ASG) has  a non-trivial central extension $c_L$, usually proportional to the angular momentum of the black hole in question. The fact that the phase space has Virasoro symmetry implies that gravity near the horizon is described by a conformal field theory with central charge $c_L$. Unlike in  most known examples, the Virasoro symmetry does not enhance an existing $SL(2,\mathbb{R})$ global conformal symmetry - as e.g. in $AdS_3$  \cite{brown} - but only a $U(1)$ isometry, given by $L_0 = \p_\varphi$. The black hole itself corresponds to a thermal density matrix in the CFT characterised by the Frolov-Thorne  temperature $T_{FT}$ \cite{frolthorne}, which becomes a purely left-moving temperature in the extremal limit. Assuming unitarity, the Cardy entropy in the CFT precisely equals the Bekenstein-Hawking entropy of the black hole

\be
S_{Cardy} = \frac{\pi^2}{3} \, c_L \, T_{FT} = S_{B.H.}
\ee

\noindent Despite the robustness of the asymptotic symmetry group computation, the Kerr/CFT holographic dictionary between fields and operators is quite poorly understood.  It is believed that the single copy of the Virasoro algebra seen from the ASG analysis is part of a larger conformal symmetry, that of a $CFT_2$, and that the black hole corresponds to a state in the DLCQ of the latter \cite{deboer}. Evidence for the $CFT_2$ interpretation as well as a rudimentary holographic dictionary were presented in the scattering amplitude computations of \cite{kerrscatt,Hartman:2009nz,Cvetic:2009jn,ChenChu} and three-point function match of \cite{Becker:2010jj}, but the exact nature of the ``$CFT_2$'' is still far from being understood.  For example, operators in this ``CFT'' have the peculiar feature that their conjectured conformal dimensions depend on the quantized angular momentum $\kappa$ of the dual spacetime field along the direction $\varphi$

\be
\Delta = a_1 + \sqrt{a_2 + a_3\, \kappa^2} \label{confdkerr}
\ee
The coefficients $a_i$ are usually only known numerically, and in the case of $4d$ Kerr for example, $a_3$ is negative and the square root  becomes imaginary for certain values of $\k$ \cite{Dias:2009ex}. This is certainly not a feature that one expects to encounter in a usual unitary CFT. Also, it is not at all understood which components of the metric in the asymptotic radial expansion should correspond to a source/expectation value for the stress energy tensor of the dual CFT, and thus there is no first-principles derivation of the relevant boundary conditions and the associated asymptotic symmetry group.

Besides the detailed holographic dictionary, there are many other interesting and important questions to address. Does the $CFT_2$ description survive also for non-extremal black holes? Hints in this direction have been provided in the works of \cite{hidden,cvetichidden}. Is the dual theory a CFT for all values of $N$ - as is the case for usual  AdS/CFT - or does it only share certain features of a CFT in the classical gravity regime (large $N$)? Is the dual theory local? Clearly, these questions are extremely relevant for understanding the microscopic description of black holes in the most general case, including the long sought-for Schwarzschild black hole. 

\bigskip

Another recent line of research that has been attempting to make contact with the real world - albeit from an entirely different perspective - is the AdS/cold atom correspondence \cite{son,balag}. Despite the name, the gravitational background relevant for this correspondence is not AdS, but rather a spacetime known as Schr\"{o}dinger, with metric

\be
ds^2 = - b^2 r^2 dt^2 + \frac{dr^2}{r^2} + 2 r dt dy + r dx^i dx^i  \;, \;\;\;\;\; i = 1, \ldots , d-2
\label{schint}
\ee
The parameter $b$ can be set to one by an appropriate rescaling of the coordinates, but we will prefer to keep it explicit. When $b=0$ we recover $AdS_{d+1}$ in Poincar\'e coordinates, but when $b \neq 0$
the asymptotics of this spacetime are radically different from those of AdS. The reason this spacetime is interesting is that it geometrically realizes the group of nonrelativistic conformal symmetries in $d-2$ spatial dimensions, also known as the Schr\"{o}dinger group. For this reason, it has been proposed
 that gravity in these spacetimes describes strongly coupled nonrelativistic CFTs in $d-2$ spatial dimensions, provided that the null direction $y$ is compactified.  
 
The holographic dictionary for these spacetimes seems to work quite similarly to the AdS case \cite{son, balag,goldberger}. For example, a massive scalar field in the bulk corresponds to an operator of nonrelativistic conformal scaling dimension
 
\be
\Delta = \frac{d}{2} + \sqrt{m^2 + \frac{d^2}{4} + b^2 \k^2} \label{confdsch}
\ee
where $\k$ is the momentum along the null $y$ direction. If $y$ is compact, then $\k$ is quantized.
 One cannot help not noticing the striking similarity between the conformal dimensions\footnote{The constants that appear inside \eqref{confdsch} can also be made arbitrarily complicated by considering less symmetric reductions along e.g. Sasaki Einstein spaces \cite{sean,bobev, donos1, donos2}. } \eqref{confdkerr} and \eqref{confdsch}, as well as between the asymptotic behaviour of \eqref{kerrm} and \eqref{schint} for $d=2$ ! 

Despite this pronounced resemblance, our way of understanding non-relativistic CFTs is quite different from our approach to Kerr/CFT. Namely, these theories are obtained by deforming the CFT dual to AdS by an operator which is \emph{irrelevant} with respect to the relativistic conformal group, but is however \emph{exactly marginal} with respect to the nonrelativistic Schr\"{o}dinger scaling symmetry \cite{nr}. Then, one  performs a DLCQ on the resulting theory, which corresponds to compactifying the null direction $y$. Interestingly \cite{kraus}, in many examples, in the supergravity limit corresponding to large $N$ and strong coupling, the deformation by a \emph{single} Schr\"{o}dinger-invariant operator becomes exact, up to issues involving multitrace operators. For the $d=2$ case we study in this paper, the corresponding deforming operators have dimension $(1,2)$. Before the DLCQ, the deformed theory is nonlocal, a fact which can be noticed from the structure of the counterterms needed in holographic renormalization \cite{nr}.

Another way to study the nonlocality of the dual theory in the $y$ direction is
to use a string-theoretical construction of certain Schr\"{o}dinger backgrounds
by use of a TsT (T-duality, shift, T-duality) transformation
\cite{Bergman:2001rw, alishganor}. This construction yields an example of a concrete field theory dual to the Schr\"{o}dinger backgrounds, known as a lightlike dipole theory \cite{berggan,alishganor}. These theories   are nonlocal deformations in the $y$ direction of the original gauge theories dual to the undeformed backgrounds. Interestingly, planar diagrams in these theories are identical to those in their undeformed counterparts \cite{noncommdip}.

\bigskip

In this paper we develop a common framework for these two apparently distinct subjects. We work with a very specific example which is more amenable to concrete computations, but we do expect our conclusions to be easily generalized. More specifically, we study solutions of type IIB supergravity on K3 which are deformations of $AdS_3 \times S^3$ and preserve  $SL(2,\mathbb{R}) \times SU(2)\times U(1)^2$ isometry. Geometrically, they correspond to (not necessarily direct) products of self-dual spacelike warped $AdS_3$ and a deformed three-sphere, where all the warping parameters are constant.  One of these backgrounds is the uplift to six dimensions of the near-horizon geometry of the extremal non-supersymmetric D1-D5-p black hole, to which a Kerr/CFT - like description applies \cite{microkerr}. All the remaining ones can be obtained by applying TsT transformations to $AdS_3 \times S^3$ followed by a U-duality, and thus are described by DLCQs of two-dimensional dipole theories and their S-duals. 

Given the high degree of resemblance between the supergravity solutions that describe the ``Kerr'' background and the dipole ones, it is natural to treat them in a unified framework. Consequently, we find that the ``Kerr CFT''  can be thought of as a nonrelativistic CFT at finite right-moving temperature, whereas the spacetimes dual to $2d$ dipole theories at finite right-moving temperature allow for a Virasoro asymptotic symmetry group very similar to that of Kerr/CFT. Consequently, $2d$ dipole theories - at least at large $N$ - should have an effective description in terms of a CFT \`a la Kerr/CFT.  A more thorough analysis of the properties and asymptotic symmetries of the spacetimes dual to $2d$ dipole theories can be found in \cite{stromdipwei}.

We have nevertheless still to define what we mean by a nonrelativistic CFT. The prescription we propose is meant as an effective definition at large $N$ and strong coupling. We define the nonrelativistic CFT as the theory one obtains by deforming the original CFT action by a \emph{single} $(1,2)$ operator which is exactly marginal with respect to non-relativistic Schr\"{o}dinger scaling

\be
S_{NR-CFT} = S_{CFT} + b \int dt dy \,\O_{(1,2)} \label{defnrcft}
\ee
All correlation functions in the deformed theory are to be computed from the ones in the original CFT by use of conformal perturbation theory. That the deformed theory makes sense without having to add any
extra operators is a major assumption which needs careful investigation. This assumption is based on the structure and consistency of the dual gravity solution.


\bigskip

The paper is structured as follows. In section \ref{bcksec} we present the backgrounds we will be studying, and classify them with respect to the symmetries they preserve and which of the 26 type IIB three-forms are turned on. We find 22 $SL(2,\mathbb{R})_L \times SU(2)_R$ invariant geometries (LR)- which we classify as self-dual (only $F_3^+$ is turned on) and dipole - and  four $SL(2,\mathbb{R})_L \times SU(2)_L$ invariant geometries (LL), which are all of dipole type. The LL backgrounds are supersymmetric, whereas the LR ones are not. The solutions have an extra global parameter $T$ which characterizes the quotient of spacelike warped $AdS_3$, and which can be roughly identified with the dimensionless right-moving temperature in the dual field theory. In section \ref{asg} we perform the ASG analysis for both types of backgrounds and find a Virasoro algebra with central charge equal to that of the undeformed $AdS_3 \times S^3$. Consequently, both should be described by CFTs in the Kerr/CFT sense.

For small values of the deformation parameters, the backgrounds can be regarded as infinitesimal 
deformations around $AdS_3 \times S^3$ that can be studied using the usual $AdS_3$/CFT$_2$ dictionary. We perform this analysis in section \ref{sugra} and find that at infinitesimal order there is a source  for a $(1,2)$ operator, which can be identified uniquely by its quantum numbers. In section \ref{3dsch} we show that there exists a ``zero-temperature'' $T \r 0$ limit of our backgrounds in which they all reduce to three-dimensional Schr\"{o}dinger spacetimes times a three-sphere. Consequently, we use our understanding of $3d$ Schr\"{o}dinger backgrounds as exact\footnote{By ``exact'' we mean that the theory is defined by the addition of a single $(1,2)$ operator, as in \eqref{defnrcft}.} deformations of a CFT (in our case, the D1-D5 CFT) by a Schr\"{o}dinger-invariant operator ($(1,2)$ in our case) to argue in section \ref{exact} that also our backgrounds correspond to exact deformations of the (DLCQ of the) D1-D5 CFT by the specific $(1,2)$ operators we had found, except that the theory is at finite right-moving temperature. 

Interestingly, all the $(1,2)$ operators we find preserve some supersymmetries: $(0,4)$ Poincar\'e supersymmetries for the LR deformation and $(4,0)$ superconformal symmetry for the LL one. The reason that the LR backgrounds are non-supersymmetric overall is due to the finite right-moving temperature $T$.
Nevertheless, given that the operator deformation is exact, the backgrounds with $T =0$ (i.e. the corresponding $3d$ Schr\"{o}dinger backgrounds), should have the same amount of supersymmetry as the operator. We check in  appendix \ref{killsp} that this is indeed the case. To our knowledge, the $3d$ Schr\"{o}dinger solutions of type IIB supergravity on K3 with eight Killing spinors were not known before.

We continue in section \ref{freeorb} with an analysis of the deforming operators from the point of view of the D1-D5 CFT. Using the map constructed in \cite{marika} between supergravity and free orbifold CFT operators, we write down very explicit expressions for the $(1,2)$ deforming operators in terms of a basis of operators at the free orbifold point of the CFT. Thus, the computation of correlators using \eqref{defnrcft} can be made very concrete. We also discuss some subtleties in the identification of the operators due to the nontrivial curvature of the moduli space \cite{kyriakos}. 

Finally, in the last two sections we discuss the applications of this analysis to the Kerr/CFT correspondence and two-dimensional dipole theories. In section \ref{d1d5rev} we first review and slightly generalize the way that the self-dual background appears in the $6d$ uplift of the near-horizon of the $5d$ Kerr-Newman black hole.  Precisely this setup has been proposed in \cite{microkerr} as an appropriate starting point for the understanding of the Kerr/CFT correspondence using string theory. In section \ref{holoint} we clarify how the point of view in that article relates to the one in the present paper. In section \ref{mxlim} we show explicitly how $3d$ Schr\"{o}dinger backgrounds appear in the near-horizon limit of this Kerr-Newman black hole, thus pointing out the relationship of Kerr/CFT to nonrelativistic CFTs. 

In section \ref{dip} we discuss the lessons we learn by using the TsT transformation to understand the dipole backgrounds. We  first review general facts about dipole theories and their connection to Schr\"{o}dinger spacetimes. Next, we show how our dipole backgrounds arise as DLCQs of both spacelike and lightlike dipole theories, and comment on how the two descriptions are related. What we learn is that   $T$ should indeed be identified with the dimensionless temperature in the dual field theory, that the DLCQ commutes with the deformation and that the holographic dictionary in asymptotically Schr\"{o}dinger spacetimes is very peculiar, once one starts considering finite temperature states. 

We end with a summary and non-technical discussion (section \ref{what}) of how our understanding of Kerr/CFT has evolved, in view of \cite{microkerr} and the present article.

\section{The backgrounds \label{bcksec}}

The supergravity backgrounds that we are interested in studying are deformations of self-dual $AdS_3 \times S^3$
 which preserve an $SL(2,\mathbb{R}) \times U(1)_{spacelike}$ subgroup of the  $AdS_3$ isometry group and an
$SU(2) \times U(1)$ subgroup of the original $S^3$ isometries. These backgrounds constitute a simple example 
  of non-AdS$_3$ geometries that can appear in the near-horizon limit of extremal black holes
 and their study might elucidate the nature of the microscopic side of the Kerr/CFT correspondence, as was 
suggested in \cite{microkerr}.

In order to have a more concrete understanding of these spacetimes, we embed them in type IIB string theory
compactified on K3. Also, we find it instructive to give a brief review of the particular type of locally
$AdS_3$ geometries - namely, self-dual $AdS_3$ - that the backgrounds of interest represent a deformation of. 
Our discussion is a summary of the results presented in \cite{deboer, pinch}.

\subsection{Self-dual and pinching orbifolds of $AdS_3$ \label{sdporb}}

The geometries that we would like to study are all deformations of the so-called spacelike self-dual
 orbifold of $AdS_3$ \cite{cousshen}. This spacetime is locally $AdS_3$ and can be obtained by 
quotienting global $AdS_3$ by a diagonal element of the $SL(2,\mathbb{R})_R$ isometry group\footnote{
An element  of the $AdS_3$ isometry group which lies entirely inside one of the $SL(2,\mathbb{R})$ factors 
is also known as a self-dual isometry generator, hence the name.}, which reduces the symmetries of this spacetime
to  only $SL(2,\mathbb{R})_L \times U(1)_R$. Its metric can be written as a $U(1)$ Hopf 
fibration over $AdS_2$

\be
ds^2 = \frac{\ell^2}{4}\left[ - \frac{r^2 d\tt^2}{T^2} + \frac{dr^2}{r^2} + \left( T d\ty + \frac{r d\tt}{T} \right)^2 \right]\;, \;\;\;\;\;
\ty \sim \ty + 2 \pi \label{sdads}
\ee
The boundary of this spacetime is located at $r \r \infty$. The coordinate $\ty$, although spacelike in the 
interior, becomes null as $r \r \infty$, and thus the boundary of this spacetime is a null cylinder. This 
geometry appears naturally in the near-horizon limit of the extremal BTZ black hole, whose full
 geometry reads

\be
ds^2 = - \frac{(\rho^2 - \rho_+^2)^2}{\rho^2} d\tau^2 + \frac{\ell^2\rho^2 d\rho^2}{(\rho^2 - \rho_+^2)^2} + 
 \rho^2 \left(R\, d\varphi - \frac{\rho_+^2 d\tau}{\rho^2}\right)^2  \label{metextrbtz}
\ee
Here $\rho_+$ is the horizon radius, $\ell$ is the $AdS_3$ length and  $\varphi \sim \varphi+2\pi$.
The extreme BTZ black hole corresponds to a thermal ensemble in a dual $CFT_2$, characterised by the following  left/right-moving temperatures\footnote{The left/right moving inverse ``temperatures'' in the CFT are linear combinations of the inverse Hawking temperature and the angular potential, and they couple to the left/right moving energy in the CFT. All the backgrounds considered in this paper have vanishing Hawking temperature, nevertheless their ``right-moving temperature'' as defined above is non-vanishing.} \cite{strexcl}
 
\be
T_L =0 \;, \;\;\;\;\; T_R = \frac{\rho_+}{\pi \ell }
\ee
Note that only the right-movers are excited, while the left-movers are in their ground state. 
The parameter $R$ appearing in \eqref{metextrbtz} is the radius of the circle on which the CFT is defined. Taking the near-horizon 
limit 

\be
\e \r 0 \;,\;\;\;\; \mbox{with}\;\; r = \frac{\rho^2-\rho_+^2}{\e}\;, \;\;\;\;\; \tt = \frac{4 R\, \e \,\tau}{\ell^2} \;, 
\;\;\;\;\; 
\ty = \varphi - \frac{\tau}{R} \label{nearhorlim}
\ee
fixed, we obtain precisely the metric \eqref{sdads} with

\be
 T =2 \pi R\,  T_R = \frac{2 R \rho_+}{\ell}
\ee
We will oftentimes be referring to $T$ as the dimensionless right-moving temperature.

The process of taking the near-horizon limit, as  explained in \cite{Strominger:1998yg,deboer,baladlcq}, amounts to performing a boost on the fixed $\rho$ sections of the extremal BTZ geometry, which becomes infinite as $\e \r 0$. This infinite boost turns the boundary cylinder of $AdS_3$ into a null cylinder. The interpretation of this process in the dual field theory is that of performing a DLCQ of the original CFT, where one keeps the entire right-moving sector while sending  the mass gap in the left-moving sector\footnote{Our conventions for what is left and what is right are inverted with respect to the earlier literature.} to infinity. Thus, in this limit, only a `chiral half' of the original CFT excitations are accessible. In spacetime, this fact is reflected in that the asymptotic symmetry group (ASG) of self-dual $AdS_3$ consists of only one, rather than two, copies of a Virasoro algebra. For the background \eqref{sdads}, written more 
simply as 

\be
ds^2 = \frac{\ell^2}{4} \left( \frac{dr^2}{r^2} + 2 r d\tt d\ty + T^2 d\ty^2 \right) \;, \;\;\;\;\; \ty \sim \ty + 2\pi
\ee
the relevant boundary conditions are \cite{deboer} 
\be
g_{\tt\tt} = \O(r^{-1}) \;, \;\;\;\;  g_{\tt r} = \O(r^{-2}) \;, \;\;\;\;\; g_{\tt \ty} =  \frac{\ell^2}{4} r +
 \O(1)\;, \;\;\;\;\;
g_{\ty \ty} = \O(1) \non
\ee

\be
g_{rr} = \frac{\ell^2}{4 r^2} + \O(r^{-3}) \;, \;\;\;\;\; g_{\ty r} = \O(r^{-1}) \label{bc}
\ee
The asymptotic symmetry group is generated by the following large diffeomorphisms 

\be
\tilde{L}_n = e^{i n \ty} (\p_{\ty} - i n r \p_r )
\ee
Note that the  boundary conditions \eqref{bc} also allow  the following three diffeomorphisms

\be
L_{-1} = i \p_t \;,\;\;\;\;\; L_0 = i (t \p_t - r \p_r) \;, \;\;\;\;\; L_1 = i \left[ (t^2 + r^{-2}) \p_t
- 2 r t \p_r - 2 r^{-1} \p_y \right] \label{lmgen}
\ee
and they are nothing but the $SL(2,\mathbb{R})_L$ isometries of the spacetime. Since the boundary conditions
\eqref{bc} do not allow for any fluctuations charged under the $L_i$, we conclude that they are trivial 
symmetry generators and therefore should not be part of the asymptotic symmetry group.

The Lie bracket algebra of the vector fields $\tilde{L}_n$ is the Virasoro algebra. Consequently, the
 Dirac bracket algebra of the 
corresponding generators is again a Virasoro, this time with a central extension

\be
c_R = \frac{3 \ell}{2 G_3}
\ee
In pure Einstein gravity, which has no local degrees of freedom,  the `perturbative' excitations of this 
spacetime are the so-called boundary gravitons, which correspond to large diffeomorphisms of the self-dual orbifold background. There is also an entire spectrum of thermal states allowed by the boundary conditions \eqref{bc}, corresponding to different values of the parameter $T$. 
The massless BTZ black hole also corresponds to a geometry allowed by these boundary conditions, namely 
the one with $T=0$. The $T=0$ spacetime is also called the null self-dual $AdS_3$ orbifold, as the compact coordinate $\ty$ is now null not only on the boundary, but also in the interior of the spacetime.

Throughout this paper we will prefer to use a slightly different version of the metric \eqref{sdads}, namely
one in which we absorb the dependence on the parameter $T$ in the identification of the coordinates. Defining

\be
y \equiv T \,\ty \;, \;\;\;\;\; t \equiv T^{-1} \, \tt
\ee
the self-dual $AdS_3$ metric takes the more elegant form

\be
ds^2 = \frac{\ell^2}{4} \left( -r^2 dt^2 + \frac{dr^2}{r^2} + (dy + r dt)^2\right) \;, \;\;\;\; y \sim y + 2 \pi T \label{natc}
\ee
Note however that, importantly, the asymptotic symmetry group has to be defined in terms of the unrescaled coordinates $\tt, \ty$,  rather than the rescaled ones. The reason is that geometries with different values of $T$, when written in terms of $y,t$ coordinates, now have a different boundary  (because the identification of $y$ changes) and cannot be simultaneously captured by a set of boundary conditions which only involve the metric components. The analogue of \eqref{bc} for the metric \eqref{natc} would only allow for the boundary gravitons, and thus will not capture all the states surviving
the DLCQ. 

Taking the $T \r 0$ limit of \eqref{natc} yields the so-called ``pinching orbifold'' of $AdS_3$. As explained
in \cite{pinch}, this limit corresponds to a low-energy limit in the dual CFT where the mass gap in both
the left-moving and right-moving sector is taken to infinity, thus leaving one with only the ground state. For this reason, the dual interpretation of such a spacetime is not very interesting, as the asymptotic symmetry group is trivial.

\subsection{Framework and notation}

Consider type IIB supergravity on $K_3$. At low energies, this yields $6d$ $4b$
supergravity \cite{Townsend:1983xt} coupled to $n=21$ tensor multiplets. The bosonic field content of this theory consists of the
graviton, five self-dual three-forms $H_3^{i+}$, $n$ anti-self-dual two-form fields $H_3^{r-}$ and $5n$ massless
scalars, which parametrise the coset\footnote{Should we wish to study the $T^4$ compactification of type IIB,
 then we should take $n=5$ instead.}

\be
\frac{SO(5,n)}{SO(5) \times SO(n)} \;, \;\;\;\;\; n=21
\ee
The ten-dimensional origin of the three-form fields is as follows: two of the self-dual forms and two of the 
anti-self-dual forms correspond to the self-dual and anti-self-dual parts of the NS-NS three-form $H_3$ and
the RR three-form $F_3$. The remaining three-form fields come from the reduction of the self-dual RR five-form
 $F_5^+$ on  the holomorphic two-cycles of K3

\be
F_5^+ = \sum_I F_3^{A+} \wedge \om^+_A + \sum_{\hat I} F_3^{\hat{S}-} \wedge \om^-_{\hat S}   
\ee
Here $\om^+_A$ represent the three self-dual two-forms on K3, while $\om^-_{\hat S}$ represent the 19 anti-self-dual
 ones of $(1,1)$ type. The correlation between the six-dimensional self-duality of the 
three-forms and that of the two-forms on K3 is due to the self-duality of $F_5^+$ in ten dimensions. 

This theory has an $AdS_3 \times S^3$ solution with radius $\ell$, in which the only nonzero field besides
 the metric is one of the self-dual three-forms in the gravity multiplet. We choose this three-form to 
be the self-dual part of the RR three-form field strength $F_3$

\be
ds^2 = ds_{AdS_3}^2 + ds_{S^3}^2\;, \;\;\;\;\; F_3^{+(0)} = \frac{2}{\ell}  (\om_{AdS_3} + \om_{S^3}) \label{ads3soln}
\ee
This background breaks the $SO(5,n)$ symmetry of the theory down to $SO(4,n)$. The moduli space of the solution is reduced to

\be
\frac{SO(4,n)}{SO(4) \times SO(n)} \;, \;\;\;\;\; n=21
\ee
We choose the $AdS_3$ metric in \eqref{ads3soln} to be that on the self-dual orbifold \eqref{natc}. This metric can be written  
 in a manifestly $SL(2,\mathbb{R})_L$-invariant fashion 
by making use of the left-invariant one-forms on $AdS_3$ (one-forms whose Lie derivative along 
the  $SL(2,\mathbb{R})_L$ vector fields \eqref{lmgen} vanishes):

\be
ds^2_{AdS_3} = \frac{\ell^2}{4}  \bigl( - w_+ w_- +  w_3^2 \bigr) 
\ee

\be
w_+ =- e^{-y} \left(\frac{dr}{r} + r dt\right)\;, \;\;\;\;\; w_-= e^y \left(\frac{dr}{r}-r dt \right) \;, \;\;\;\;\; w_3 = dy + rdt\;, \;\;\;\;\; 
\ee
Similarly, the metric on $S^3$, which has isometry group $SU(2)_L \times SU(2)_R$, can be written in a 
manifestly $SU(2)_R$-invariant fashion 
by using the following right-invariant one forms on $S^3$: 
\be
\s_1 =\cos \psi d\th + \sin \th \sin \psi d\phi \;, \;\;\;\;\; \s_2 = -\sin \psi d \th +
 \sin \th \cos \psi d\phi
\ee

\be
 \s_3 = d \psi + \cos{\th} d \phi\;, \;\;\;\;\;ds^2_{S^3} =\frac{ \ell^2}{4} (\s_1^2 + \s_2^2 + \s_3^2)
\ee
Should we want to write the $S^3$ metric in a manifestly $SU(2)_L$-invariant fashion instead, we should then use the 
left-invariant one forms\footnote{So far, the labeling of isometries as left-moving versus right-moving is just a matter
 of convention, but the relative orientation  does 
become meaningful once we consider supersymmetry. }

\be
\s_{1L} =\cos \phi d\th + \sin \th \sin \phi d\psi \;, \;\;\;\;\; \s_{2L} = -\sin \phi d \th +
 \sin \th \cos \phi d\psi
\ee

\be \s_{3L} = d \phi + \cos{\th} d \psi \;, \;\;\;\;\; ds^2 = \frac{\ell^2}{4} \sum_i \s_{iL}^2
\ee
Note that there is a symmetry between the $AdS_3$ and the $S^3$ factors, in the sense that they are both 
written as Hopf fibrations over a lower-dimensional Einstein space. 
\bigskip

To summarize, the undeformed background that we start from is the $AdS_3 \times S^3 $ solution of type IIB
supergravity on $K3$

\be
ds^2 = ds_{AdS_3}^2 + ds_{S^3}^2=  \frac{\ell^2}{4} \bigl( - w_+ w_- +  w_3^2 + \s_1^2 + \s_2^2 + \s_3^2 \bigr)
\label{bkgnd}
\ee
supported by RR 3-form flux

\be
F_3^{(0)} = \frac{2}{\ell} (\omega_{AdS_3} + \omega_{S^3})=\frac{\ell^2}{4} \bigl( \s_1 \wedge \s_2 \wedge \s_3  + \half w_+ 
\wedge w_- \wedge w_3\bigr) \;, \;\;\;\; F_3^{(0)} = \star_6 F_3^{(0)} \label{bkf}
\ee
If we ignored the identification of the $y$ coordinate for a moment, the above background would 
 preserve sixteen supersymmetries. We will find it convenient to split the ten-dimensional Killing spinors into a six-dimensional and a four-dimensional, internal, part. Four of the corresponding six-dimensional  Killing spinors depend explicitly only on $t, r, \psi$, whereas the other four  depend explicitly only on the remaining coordinates. Since there are two chiral, covariantly constant, spinors on K3, we have a total of 16 supersymmetries in ten dimensions. The dependence  on the coordinates is such that the first four $6d$ Killing spinors are invariant under  $SL(2,\mathbb{R})_L \times SU(2)_L$, whereas the remaining four are invariant under  $SL(2,\mathbb{R})_R \times SU(2)_R$. Thus, as expected, supersymmetry induces a preferred pairing between the $AdS_3$ and $S^3$ coordinates, and thus of the $SL(2,\mathbb{R})$ and $SU(2)$ isometry group factors. Should we want to reverse this pairing we can for example change the self-duality the three-form $F_3^{(0)}$ into anti-self-duality.

Breaking any of the sphere or $AdS_3$  isometry factors leads to broken superconformal symmetry  on the corresponding side. For example, the identification of $y$ breaks $SL(2,\mathbb{R})_R$ and thus the only preserved Killing spinors are the left-moving ones. If in addition we break $SU(2)_R$ the supersymmetries of the background are still the left-moving $(4,0)$ superconformal ones\footnote{Our terminology is such that $(4,0)$ superconformal symmetry assumes the existence of eight Killing spinors, whereas $(4,0)$ Poincar\'e or usual supersymmery assumes the existence of only four Killing spinors.}, whereas if we break $SU(2)_L$ instead no more supersymmetries are preserved.

\subsection{Backgrounds with $SL(2,\mathbb{R}) \times SU(2)$ isometry}

We would now like to classify and study solutions of type IIB supergravity which preserve 
$SL(2,\mathbb{R})_L \times SU(2) \times U(1)^2$ isometry. These solutions should be thought of as 
finite deformations of the $AdS_3 \times S^3$ background \eqref{bkf}, which is supported by purely self-dual RR $F_3$ flux. Also, for simplicity we only want to consider the case in which at most one other three-form field is turned on in addition to $F_3^{(0)}$. Given the $SO(5,21)$ symmetry which rotates the 26 three-forms of type IIB supergravity on K3 into each other, we only need to consider three cases:

\bi
\item only $F_3^+$ is turned on
\item another self-dual three form is turned on in addition to $F_3^{+(0)}$
\item an anti-self-dual three form is turned on in addition to  $F_3^{+(0)}$
\ei 
There are four different possibilities falling in the second category, and $21$ in the third. 
To study them it suffices to consider the $O(2,2)$ invariant
 consistent truncation of  type IIB on K3 put forth in \cite{duffmult}. The action reads

\bea
\mathcal{L}_{6d} &=&  R - \half (\p \phi_1)^2 - \half e^{2 \phi_1}(\p \chi_1)^2- \half (\p \phi_2)^2 -
 \half e^{2 \phi_2}(\p \chi_2)^2 -\non \\
&& \hspace{2 cm} - \frac{1}{12} e^{-\phi_1-\phi_2} H_{3}^2 -
\frac{1}{12} e^{\phi_1-\phi_2} (F_3 + \chi_1 H_3)^2 + \chi_2 H_3 \wedge F_3 \label{duffa}
\eea
where $H_3$ is the NS-NS three-form and $F_3$ is the RR one. The scalar $\phi_1$ is the ten-dimensional dilaton, whereas $e^{-\phi_2}$ represents the volume of the internal manifold in ten-dimensional Einstein frame.  The combination appearing in front of $F_3^2$ is the volume of K3 in string frame, which is an attracted scalar.

Let us first consider solutions of this theory which preserve an $SL(2,\mathbb{R})_L \times SU(2)_R \times U(1)_L 
\times U(1)_R$ isometry. The scalars have to be constant, and their equations of motion yield the constraints

\be
F_3 \wedge H_3 = F_3 \wedge \star H_3 = 0 \;, \;\;\;\;\;F_3^2=H_3^2 =0
\ee
The remaining equations of motion read

\be
R_{\mu \nu} = \frac{1}{4} e^{\phi_1-\phi_2} F_{\mu\a\b} F_\nu{}^{\a\b}+ \frac{1}{4} e^{-\phi_1-\phi_2}
 H_{\mu\a\b} H_\nu{}^{\a\b}, \;\;\;\;\;\;\; d\star F_3 = d\star H_3 =0
\ee
Note that the equations above do not fix the values of the scalar fields $\phi_{1,2}$ or the overall
normalization of the flux, namely we only obtain solutions for

\be
\hat{F}_3 = e^{\frac{\phi_1-\phi_2}{2}} F_3 \;, \;\;\;\;\; \hat{H}_3 = e^{- \frac{\phi_1+\phi_2}{2}} H_3
\ee

\bigskip

\bigskip

\noindent The first case we would like to study is the one in which only $F_3$ is nonzero and is self-dual.
We write the fields in  terms of left (right) invariant forms on $AdS_3$  ($S^3$), such that the 
isometries are manifest. In this case, the solution for the  metric and  $F_3$ reads \cite{josh}
(before imposing self-duality)

\be
ds^2 =\frac{ \ell^2}{4 h} \bigl( - w_+ w_- + \g w_3^2 + \s_1^2 + \s_2^2 + \g \s_3^2 + 2 \g \e_g w_3 \s_3 \bigr) \label{bck}
\ee

\be
\hat F_3 = \frac{\ell^2}{4} \left[ \left( \s_1 \wedge \s_2 \wedge \s_3  + \half w_+ \wedge w_- \wedge w_3\right) + \frac{ \g \,
 \e_B}{h} \left(\s_1 \wedge \s_2 \wedge w_3  + \half w_+ \wedge w_- \wedge \s_3 \right)\right] \non 
\ee
This is a two-parameter family of solutions, parametrized by $\e_g$ and $\e_B$. The remaining constants are 
given by

\be
\g = \frac{1-\e_g^2}{1+\e_B^2 - \e_g^2 - \e_B \,\e_g \sqrt{1+\e_B^2 - \e_g^2}}\;, \;\;\;\;\; h = \g \sqrt{1+\e_B^2 - \e_g^2}
\label{gamh}
\ee
Imposing $F_3 = \star F_3 $ yields the relation

\be
\e_B^2 = \half \sqrt{1-\e_g^2} \; (1-\sqrt{1-\e_g^2}) \;, \;\;\;\;\; \e_B \e_g >0
\ee
We call this background ``self-dual'', because it is created by a single self-dual flux. It is precisely it 
which arises in the near-horizon
of the uplift of the  $5d$ non-supersymmetric Kerr-Newman black hole considered in 
\cite{microkerr} and reviewed in section \ref{d1d5rev}. A convenient parametrisation of $\e_B, \e_g, \g, h$ for the 
self-dual backgrounds is in terms of a new parameter $\d$ such that

\be
\e_g = \frac{2 \cosh 2\d}{1+ \cosh^2 2 \d} \;, \;\;\;\;\; \g = 1 + \frac{1}{\cosh^22 \d} \;, \;\;\;\; 
\e_B = \frac{ \sinh 2 \d}{1+ \cosh^2 2 \d}\;, \;\;\;\;\; h = \tanh 2 \d \label{sdcond}
\ee
The solution with $\hat F^+_3 = \hat F^{(0)}_3$ and $\hat H_3 = \hat H_3^-$ can be
obtained from \eqref{bck}-\eqref{gamh} by simply solving for $\hat F^+_3 = \hat F^{(0)}_3$ (this yields $\e_g=0$) and then letting
$\hat{H}_3^-$ equal the anti-self-dual part of the solution above with $\e_g=0$. More explicitly

\be
ds^2 = \frac{ \ell^2}{4} \sqrt{1+ \e_B^2} \,\bigl( - w_+ w_- + \g w_3^2 + \s_1^2 + \s_2^2 + \g \s_3^2  \bigr) \;, \;\;\;\;\;
\g= \frac{1}{1+ \e_B^2} \non
\ee

\be
\hat{H}_3^-= e^{-\frac{\phi_1 + \phi_2}{2}} H_3^- = \frac{\ell^2  \e_B}{4 \sqrt{1+\e_B^2}} \bigl(\s_1 \wedge \s_2 \wedge w_3  + \half w_+ \wedge w_- \wedge \s_3 \bigr)\;,
\;\;\;\;\; F_3 = F_3^{(0)} \label{nsdip}
\ee
The above background is nonsupersymmetric and can be obtained by performing a TsT transformation (T-duality, shift, T-duality) on the $AdS_3 \times S^3$ solution \eqref{bkgnd}. This transformation is discussed at length in section \ref{dip}. The dual field theory is a so-called dipole field theory \cite{berggan}. We therefore call these $\e_g=0$ backgrounds \emph{$SL(2,\mathbb{R})_L \times SU(2)_R$ invariant} (or simply LR) \emph{dipole} backgrounds.

The remaining case to study is when we have an extra self-dual field $H_3^+$ in the background 
given by $F^{+ (0)}$. Self-duality of $F^{(0)}$ requires that $\e_g=0$, whereas self-duality of 
 $H_3$ requires $\e_B=0$. Therefore there are no nontrivial self-dual deformations  which preserve
 $SL(2,\mathbb{R})_L \times SU(2)_R \times U(1)_L \times U(1)_R$  isometry  other than that induced by the very field which
produces the background.

The deformations with $H_3^+$ turned on can instead preserve  an $SL(2,\mathbb{R})_L \times
 SU(2)_L \times U(1)_R^2$ subgroup of the original $AdS_3 \times S^3$ isometries. This can be
easily seen from the fact that when we interchange $SU(2)_L$ with $SU(2)_R$ (i.e. we interchange
 $\psi$ and $\phi$), the self-duality condition becomes anti-self-duality and vice-versa. 
Thus, the dipole-type deformations $(\e_g=0)$  are now self-dual, and the metric and field 
strengths read

\be
ds^2 = \frac{\ell^2}{4} \sqrt{1+\e_B^2} \bigl( - w_+ w_- + \g w_3^2 + \s_{1L}^2 + \s_{2L}^2 + \g \s_{3L}^2  \bigr) \non
\ee

\be
\hat F^+_3 = \frac{\ell^2}{4} \bigl( \s_1 \wedge \s_2 \wedge \s_3  + \half w_+ \wedge w_- \wedge w_3\bigr) 
= \frac{\ell^2}{4}\bigl( -\s_{1L} \wedge \s_{2L} \wedge \s_{3L}  + \half w_+ \wedge w_- \wedge w_3\bigr) \non
\ee

\be
 \hat{H}^+_3 =\frac{\ell^2 \, \e_B}{4 \sqrt{1+\e_B^2}} \bigl(\s_{1L} \wedge \s_{2L} \wedge w_3  + \half w_+ \wedge w_- \wedge \s_{3L} \bigr)\;,
\;\;\;\;\; \g=\frac{1}{1+\e_B^2} \label{ssdip}
\ee
Since the isometry group of these backgrounds is $SL(2,\mathbb{R})_L \times SU(2)_L $, half the superconformal symmetry is preserved, namely $(4,0)$. We consequently call them \emph{supersymmetric} or $SL(2,\mathbb{R})_L \times SU(2)_L$ invariant (LL) \emph{dipole} backgrounds.

\bigskip

In conclusion, we have found three types of backgrounds: the self-dual background ($F_3 = \star 
F_3$), the supersymmetric (LL) dipole backgrounds ($F_3 = F_3^{(0)}$ and  $H_3^+ \neq 0 $)
and the LR dipole backgrounds ($F_3 = F_3^{(0)}$ and  $H_3^- \neq 0$).
The remaining solutions of type IIB supergravity  respecting the same isometries can be obtained 
by applying $SO(4) \in SO(4,21)$ transformations to the LL dipole backgrounds or $SO(21) \in SO(4,21)$
transformations to the LR ones. These transformations do not change the background metric, 
the supporting self-dual RR field or the supersymmetry properties of the solution, but do rotate the remaining
 three-forms into each other. The resulting backgrounds therefore have $F^{(0)}_3$ and one of
  $F_3^{A+}$ turned on in the supersymmetric LL case, and $F_3^{(0)}$ and one of $F_3^{\hat S -}$, $F_3^-$ 
turned on in the nonsupersymmetric LR case. The dual field theories to these backgrounds could therefore 
be called ``S-duals'' of dipole theories, because they are related by a $SO(4,21)$ transformation to the usual dipole 
backgrounds. For example, a deformation which has $F_3^{\hat S -}$ turned on can be obtained by applying
 a sequence of an S-duality, a mirror symmetry transformation on K3 and an S-duality back on the
standard nonsupersymmetric dipole deformation. In the rest of the paper we will oftentimes denote the  dipole backgrounds and their S-duals by the same name, and hope this will not cause much confusion. 

As far as the metric is concerned, all the above backgrounds are products (not necessarily direct) of a quotient of spacelike warped\footnote{Spacelike warped $AdS_3$ is a $U(1)$ Hopf fibration over $AdS_2$ very similar to \eqref{sdads}, but where the fibre is either stretched or squashed by a constant factor.} $AdS_3$ and a deformed sphere. Quotients of spacelike warped $AdS3$ have recently received much attention in the literature \cite{stromgirls,hopfpuff,Moussa:2008sj}, 
and it has been shown that in certain cases they can be interpreted as a temperature in the dual field theory. In our case, the quotient should represent a purely right-moving temperature. 

The supersymmetry or lack thereof of the deformed backgrounds can be easily understood: as we have already mentioned, four of  the Killing spinors of unquotiented $AdS_3 \times S^3$ are $SL(2,\mathbb{R})_R \times SU(2)_R$ invariant, while the remaining four are $SL(2,\mathbb{R})_L \times SU(2)_L$ invariant. In ten dimensions the amount of supersymmetry is doubled. Thus, the backgrounds that preserve the conformal and the spherical symmetry on the same side should have eight superconformal symmetries in $10 d$. The backgrounds that preserve the $SL(2,\mathbb{R})$ and $SU(2)$ factors on opposite sides may preserve the supersymmetries associated with the $SU(2)$ factor, but not the superconformal symmetries on the corresponding side. Nevertheless, since the $y$ quotient induces a temperature on the side where supersymmetry could have been preserved, we do not expect the $SL(2,\mathbb{R})_L \times SU(2)_R$ invariant backgrounds to be supersymmetric at all.

\subsection{Asymptotic symmetry groups \label{asg}}

In this section we would like to show that the backgrounds \eqref{bck} admit boundary conditions which yield an asymptotic symmetry group (ASG) that consists of a centrally-extended Virasoro algebra. This fact strongly suggests that the dual description of our backgrounds should be in terms of a (chiral half of a) CFT. Interestingly, the central charge of the putative dual CFT is \emph{the same} as that of the original $AdS_3$ background we have been deforming, provided the $F_3$ flux through the $S^3$ is kept fixed as the deformation is turned on. 

We consider the following boundary conditions on the metric and on the two-form potentials
\begin{flushleft}
\vspace{- .7 cm}
\be
\d g_{\mu\nu} = \O\left(\begin{array}{cccccc} r^2 & r^{-2} & 1 & r^{-1} & r & r\\ & r^{-3} & r^{-1} & r^{-2} & r^{-2} & r^{-2} \\ & & 1 & r^{-1} & 1 & 1 \\ & & & r^{-1} & r^{-1} & r^{-1}\\ & & & & r^{-1} &  r^{-1} \\ & & & & & r^{-2} \end{array} \right)\;\;\;\;\;\;\;\;\;
\d B_{\mu\nu} = \O\left(\begin{array}{cccccc} 0 & r^{-2} & r^{-1} & r^{-1} & r & r\\ & 0 & r^{-3} & r^{-3} & r^{-2} & r^{-2} \\ & & 0 & r^{-2} & r^{-1} & r^{-1} \\ & & & 0 & 1 & r^{-1}\\ & & & & 0& r^{-1}  \\ & & & & & 0 \end{array} \right) \non
\ee
\end{flushleft} 
\vspace{- .3 cm}
\be  \label{bndcond} \ee
in the basis $(t,r,y,\th,\phi,\psi)$. These boundary conditions are left invariant  by the following diffeomorphisms

\be
\xi_f = f(y) \p_y - r f'(y) \p_r 
\ee
which should be simultaneously accompanied by the following gauge transformation

\be
\d \L = - \frac{\g \e_B}{h} \,f(y) \, \s_3
\ee
The Lie bracket algebra of these vector fields is 
\be
[\xi_f, \xi_g]_{L.B.} = \xi_{(f g' - f' g)}
\ee
Expanding the above vector fields $\xi_f(y)$ in Fourier modes $\xi(y) = \sum_n \xi_n e^{i n \frac{y}{T}}$, we find that the above Lie bracket algebra is the Virasoro algebra. The Dirac bracket algebra of the associated generators is 

\be
\{\mathcal{Q}_{\xi_f} , \mathcal{Q}_{\xi_g}\}_{D.B.} = \mathcal{Q}_{\xi_{(f g' - f' g)}} + \int_{\p \mathcal{M}} K_{\xi_f} (\mathcal{L}_{\xi_g} \bar \Phi, \bar \Phi) \label{asgg}
\ee
Here $\bar \Phi$ denote the background metric and two-form fields in the theory, and the expression for $K_\xi$ can be found, including the contribution of the two-form potentials, in \cite{compform}. The integral is performed over the spatial boundary of the spacetimes, which in our case is $S^3 \times S^1_y$. The last term in the equation above is non-vanishing and yields a central extension to the Virasoro asymptotic symmetry group\footnote{The expression below only encompasses the gravitational contribution to the central charge. In principle, we should also check that the two-form potentials do not contribute. Nevertheless, it has been shown in \cite{compext} that for a large class of geometries which are $U(1)$ fibrations over $AdS_2$ - as is the case here - the central charge of the asymptotic symmetry algebra only receives contributions from the gravitational part of the action, so we have found this extra computation unnecessary.}



\be
 c = c_0 \frac{\g \sqrt{1-\e_g^2}}{h^2} = \frac{c_0}{\sqrt{1-\e_g^2}} \left(
1-\frac{\e_B \e_g}{\sqrt{1+\e_B^2 -\e_g^2}} \right) 
\ee
where $c_0$ is the central charge for the $AdS_3 \times S^3$ solution with flux $\ell^2/4$

\be
c_0 = \frac{3 \pi^2 \ell^4}{G_6} = \frac{3 \ell}{ 2 G_3}
\ee
in which we recognize the well-known Brown-Henneaux result. For the dipole backgrounds

\be
\e_g =0 \;, \;\;\;\;\; h = \sqrt{\g} \;\;\; \Rightarrow \;\;\; c=c_0
\ee
For the self-dual (Kerr) background 

\be
\e_g = \frac{2 \cosh 2 \d}{1+ \cosh^2 2 \d} \;, \;\;\;\;\; \g = 1 + \frac{1}{\cosh^2 2\d} \;, \;\;\;\;\; h = \tanh 2 \d \;\;\;
\Rightarrow \;\;\; c = c_0
\ee
We thus find the very interesting fact that, even though the radius of the geometry changes from $\ell$ to $\ell/\sqrt{h}$ upon turning on the deformation, the central extension of the asymptotic symmetry group is unaffected, as long as we perform a single deformation at a time.

Another interesting observation is that the product

\be
S_{Cardy} = \frac{ \pi^2 c}{3}  R\, T_R \label{cardy}
\ee
which represents a putative Cardy entropy in the dual field theory - if one takes the existence of the Virasoro asymptotic symmetry algebra to its full implications and interprets $R\, T_R$ as the dimensionless right-moving temperature - is unchanged from the $AdS_3$ case.  In the dipole theory, this can be understood from the fact that at large N diagrams with no external legs are not affected by the perturbation, and so the entropy is not modified. It would be very interesting to have a similar interpretation for the self-dual backgrounds and thus for Kerr/CFT.

\section{Supergravity analysis of the perturbation \label{sugra}}

All the backgrounds we have presented so far, for infinitesimal values of the deformation parameter, can 
be thought of as small deformations around self-dual $AdS_3 \times S^3$ and thus be interpreted using the usual 
AdS/CFT dictionary. To do this, we first need to reduce our $6d$ supergravity solutions on the $S^3$.
Fortunately, the reduction and identification of the dictionary has been performed in great detail in
\cite{dkss} for the entire linearized spectrum, so we only have to pick out the results we need. The only slight subtlety in our analysis is that we are studying perturbations around thermal AdS, whose interpretation  is slightly different from what Poincar\'{e}-coordinates intuition would lead one to think.

\subsection{Holography in a thermal background} \label{sugran}

The background metric is $AdS_3 \times S^3$ in coordinates

\be
\frac{ds^2}{\ell^2} = \frac{\ell^2}{4} \left(  -r^2 dt^2 + \frac{dr^2}{r^2} + (dy + rdt)^2 \right) + \ell^2 d \Omega^2_3
\ee
Now we expand \eqref{bck} to first order in $\e = \e_B$ and/or $\e_g$, and interpret the change in the metric and the two-form potentials $B^{(2) I}$ as a linearized perturbation around $AdS_3 \times S^3$, which has a well-defined holographic interpretation. Since

\be
\g = 1 + \O(\e^2) \;, \;\;\;\;\; \e_{B}\sim \e_g = \O(\e)
\ee
at linearized order the $SL(2,\mathbb{R})_L \times SU(2)_R$ - invariant perturbations we are interested in studying correspond to a mixed component of the $6d$ metric and of the $6d$ B-fields, of the form

\be
\Delta g_{\mu a} \propto  \Delta B_{\mu a}^I \propto (dy+rdt)(d\psi + \cos\th d\phi) \label{formpert}
\ee
whereas the $SL(2,\mathbb{R})_L \times SU(2)_L$ - invariant perturbations correspond to a mixed component of the $B$ fields only

\be
\Delta B_{\mu a}^I \propto (dy+rdt)(d\phi + \cos\th d\psi) 
\ee
Here $x^{\mu}$ represent the coordinates on $AdS_3$, while $y^a$ represent the coordinates on $S^3$. 
Such mixed components yield massive KK vector fields in $3d$, whose mass is determined by the $S^3$ quantum number $l$. The relevant equations of motion take the form \cite{dkss}

\be
F_{\mu\nu} + \frac{\l}{\ell} \, \e_{\mu\nu\rho} A^\rho =0 \;, \;\;\;\;\; \l =  \half (l+1)
\ee
and indicate that the vector field is topologically massive and only carries one degree of freedom.  
The dependence on the $S^3$ coordinates implies that in our case $l=1$, but we do not need 
this right now. We would like
to look for solutions of the above equation which are independent of $y$ and $t$. We find a 
non-normalizable solution

\be
A_s = r^\l dt + \frac{\l}{2\l-1} \,r^{\l-1} dy \label{asource}
\ee
as well as a normalizable one

\be
A_{v.e.} = r^{-\l} dy \label{avev}
\ee
The entire perturbation \eqref{asource} represents a constant source for an  irrelevant operator of dimension $(\l, \l +1)$, whereas \eqref{avev} represents a constant expectation value for the same operator in a state of non-zero right-moving temperature $T_R$. 

The three-dimensional part of the perturbations \eqref{formpert} precisely matches the  the non-normalizable perturbation $A_s = dy + r dt$ for $\l=1$. Thus, in the background under consideration, only a constant source for a $(1,2)$  operator is
turned on. The constant term $dy$ in $A$ does not represent, as one might have naively thought from 
experience with perturbations of Poincar\'{e} $AdS_3$, an expectation value or source for a dual current, but is simply part of the non-normalizable mode of the massive vector field $A_\mu$.

\subsection{Quantum numbers \label{qno}}

In this section we work out in detail the quantum numbers of each of the linearized perturbations we
are interested in studying. First, we expand the linearized metric and two-form field perturbations in scalar and 
vector harmonics on $S^3$. Following  \cite{dkss} we introduce the notation

\be
\Delta g_{\mu a}  = \sum K_\mu^{(\ell, \pm 1)} (x) Y_a^{(\ell, \pm 1)} (y) + K_\mu^{(\ell 0)} \p_a Y^{\ell 0}(y)  
\ee
where  $Y^{\ell 0} (y)$ are scalar harmonics on $S^3$ and $Y_a^{\ell, \pm 1}(y)$ are the corresponding vector harmonics, which satisfy

\be
\Box_y Y_a^{(\ell, \pm 1)} = [2-(\ell+1)^2] Y_a^{(\ell, \pm 1)} \;, \;\;\;\;\; \nabla^a Y_a^{(\ell,\pm 1)} =0
\ee
Similarily, we write

\be
\Delta B_{\mu a}^I  = \sum Z_\mu^{I \,(\ell, \pm 1)} (x) Y_a^{(\ell, \pm 1)} (y) + Z_\mu^{I\,(\ell 0)} \p_a Y^{\ell 0}(y)  
\ee
Here the index $I$ labels the $5+n$ three-form fields of $6d$ supergravity, with $n=21$ for the K3 compactification and $n=5$ for the $T^4$ one. The $SL(2,\mathbb{R})_L \times SU(2)_R$ invariant perturbation  \eqref{formpert} corresponds to $Y_a^{(1, 1)}$. Using the isometry between the  $SO(4)$ rotations of the sphere and $SU(2)_L \times
 SU(2)_R$, one
finds that the map between the $SO(4)$ labels $(l_1,l_2)$ in $Y^{l_1,l_2}$ and the isospins $(j_L, j_R)$ is

\be
j_L = \half (l_1 + l_2) \;, \;\;\;\;\; j_R = \half (l_1 - l_2)
\ee
So, we finally find that the $SU(2)_L \times SU(2)_R$ quantum numbers of the operator are 

\be
(j_L, j_R) = (1,0)
\ee
Since the perturbation is independent of both $\phi$ and $\psi$, its $j_{L,R}^3$ quantum numbers,
which we denote by $m_{L,R}$, are zero

\be
m_L = m_R =0 \label{mlmrz}
\ee
All perturbations we consider correspond to turning on a massive vector field in $3d$, $K_\mu \sim Z_\mu \sim A_\mu$ 

\be
A = dy + r dt
\ee
As already discussed, the radial dependence of $A_\mu$  indicates that it represents a source for an operator of conformal weights 

\be
(h_L, h_R) = (1,2) 
\ee 
According to \cite{dkss} there are $n+1$ vector fields with these quantum numbers in $AdS_3$, transforming as a singlet and a vector of $SO(n)$. The fields transforming in the vector representation of $SO(n)$ descend from the cross components of the $n$ anti-self-dual tensor fields, whereas the singlet corresponds to  one particular linear combination of the mixed metric and the self-dual $F_3^+$ field. Consequently, the LR invariant dipole deformations transform as a vector under $SO(n)$, whereas the self-dual deformation corresponds to the $SO(n)$ singlet.

We also classify the deformations preserving $SL(2,\mathbb{R})_L \times SU(2)_L$. There the corresponding
spherical harmonic turned on is $Y^{1,-1}_a$, while the AdS form of the perturbation is as before. We conclude that the perturbation corresponds to a source for an operator with quantum numbers

\be
(h_L,h_R)=(1,2) \;\;\;\; \mbox{and} \;\;\;\; (j_L,j_R)= (0,1)
\ee
in addition to \eqref{mlmrz}. There are four such operators transforming in the $(2,2)$  of $SO(4)_{outer}$.

\bigskip

Having established the $SL(2,\mathbb{R})_L \times SL(2,\mathbb{R})_R$ and $SU(2)_L \times SU(2)_R$ quantum  numbers of the deformations, we can also work out their supersymmetry properties. The spectrum of KK excitations of $6d$ supergravity around $AdS_3 \times S^3$ is organized in short representations of the $SU(1,1|2)_L \times SU(1,1|2)_R$ supergroup \cite{dkss, deboersusy}. There are four left-moving supersymmetry generators $G^{\a a}$ and four right-moving ones $\tilde{G}^{\a a}$, with $\a,a = \{ +, - \}$, obeying the reality condition $G^{\a a} = \e^{\a\b} \e^{ab} (G^{\b b})^*$ (and similarly on the right)\footnote{We are using the conventions of \cite{kyriakos}.}.The first index is the $SU(2)_{L/R}$ R-symmetry index, whereas the second one transforms under an $SU(2)_{L/R}$ outer automorphism group of the $\N=4$ algebra, which only rotates the fermionic generators while leaving the bosonic ones unchanged. The outer automorphism group $SO(4)_{outer} \cong SU(2)_L^{out} \times SU(2)_R^{out}$ is identified with the spacetime $SO(4) \subset SO(5,21)$ subgroup which rotates $H_3^+$ and $F_3^{A+}$ into each other. The $\N=4$ superconformal algebra also contains three R-symmetry currents $J_0^i = \s_{\a\b}^i J^{\a\b}_0$, which transform as a vector under $SU(2)_L$ R-symmetry.  

A short multiplet is constructed starting from a chiral primary state of weight $h \in \half \,\mathbb{Z}$, $| \chi \rangle$, 
which satisfies

\be
G_{-\half}^{++}| \chi \rangle = G_{-\half}^{+-}| \chi  \rangle =0\;, \;\;\;\;\;L_0| \chi  \rangle = J^3_0 | \chi  \rangle =
 h | \chi  \rangle \;, \;\;\;\; J^{++}_0 | \chi  \rangle =0
\ee
in addition to the usual requirements that it be a superconformal primary

\be
L_n | \chi  \rangle = J^i_n |\chi  \rangle  = G^{\a a}_{n-\half} | \chi  \rangle =0 \;, \;\;\;\; \forall \, n >0
\ee
Acting with $J^{--}_0$ lowers the spin projection on the `3' axis by one. Acting with it the maximum of $2h$ times generates a spin $h$ multiplet of $SU(2)_L$ R-symmetry. On the other hand, acting with $G^{-a}_{-\half}$ lowers the total spin while increasing the conformal dimension. Since we can act with $G^{-a}_{-\half}$ at most twice  before annihilating the state, the structure of a short supermultiplet is
\medskip

\begin{center}
\begin{tabular}{r|cc}
state & $j_L$ & $L_0$ \\ \hline
$| \chi \rangle $ & $ h $ & $h$ \\
$ G_{-\half}^{-+}| \chi  \rangle , G_{-\half}^{--} | \chi  \rangle$  & $h - \half$ & $ h+ \half$ \\
$ G_{-\half}^{-+} G_{-\half}^{--}  | \chi \rangle $ & $h-1$ & $h+1$ 
\end{tabular}
\end{center}
\medskip

\noindent  The representation content under the full $(4,4)$ supersymmetry is obtained via the tensor product of short multiplets on the left with their right-moving counterparts. In the following, we shall denote right-moving quantities by a tilde. 

Given that the $SL(2,\mathbb{R})_L \times SU(2)_R$ invariant deformations  have $h_L - j_L =0$, 
$h_R - j_R =2$ and $m_L = m_R =0$, the corresponding operators must take the form

\be
\O_{(1,2)}^{LR} = J_0^{--} \tilde{G}^{--}_{-1/2} \tilde{G}^{-+}_{-1/2} \mathcal{O}_{(1,1)}^\chi \label{nso}
\ee
where $ \mathcal{O}_{(1,1)}^\chi$ is a $(1,1)$ chiral primary operator. There are $n+1=22$ such chiral primaries, transforming as a vector and a singlet of $SO(n)$. Note that the above operators are $\tilde{G}_{-\half}$ exact, so they preserve full $(0,4)$ Poincar\'e supersymmetry. Since the operator is not marginal on the right ($h_R =2$),  the superconformal symmetries are broken. 

As far as the four $SL(2,\mathbb{R})_L \times SU(2)_L$ invariant deformations are concerned, we have
$h_L - j_L = h_R - j_R =1$, so the operators they correspond to must take the form

\be
\O_{(1,2)}^{LL} = \tilde{J}_0^{--} G_{-\half}^{-a} \tilde{G}_{-\half}^{-b} \mathcal{O}_{\left(\half,\frac{3}{2}\right)}^\chi \label{sso}
\ee
As expected, these transform in the $(2,2)$ representation of $SO(4)_{outer}$. $\mathcal{O}_{\left(\half,\frac{3}{2}\right)}^\chi $ is a chiral  primary of the indicated dimension. Note that these operators are exactly marginal on the left (they have dimension one and zero R-charge), and thus they preserve the full left-moving superconformal symmetry,  but they break the right-moving one. These operators represent the supersymmetric dipole deformations.

\subsection{Relationship to $3d$ Schr\"{o}dinger spacetimes \label{3dsch}}

Schr\"{o}dinger spacetimes are $d+1$ - dimensional spacetimes whose isometry group is the Schr\"{o}dinger group of nonrelativistic conformal transformations in $d-2$ spatial dimensions. Their metric reads

\be
ds^2 = - b^2 r^2 dt^2 + \frac{dr^2}{4 r^2} + r (2 dt dy + dx^i dx^i) \;, \;\;\;\;\; i = 1, \ldots, d-2
\label{ddimschroed}
\ee
The parameter $b$ can be set to one by appropriate rescalings of the $t$ and $y$ coordinates, but in the following we will prefer to keep it explicit. The isometries of this metric consist of translations, rotations, Galilean boosts, nonrelativistic scale transformations, under which

\be
x^i \r \l \, x^i \;, \;\;\;\;\; t \r \l^2 t\;, \;\;\;\;\; y \r y \;, \;\;\;\;\; r \r \l^{-2} r
\ee
and also nonrelativistic special conformal transformations. Due to the above invariance under nonrelativistic rescalings, these spacetimes have been proposed as holographic duals of strongly coupled nonrelativistic conformal field theories \cite{son, balag}, and have since received plenty of attention (a very incomplete list of references includes \cite{Herzog:2008wg, Maldacena:2008wh, Adams:2008wt,goldberger,Barbon:2008bg,baltbob, baltmulti, kraus, nr}).  

In the following we will specialize to three-dimensional Schr\"{o}dinger backgrounds, which geometrically realize the  nonrelativistic  conformal group in zero spatial dimensions. The isometry group is now $SL(2,\mathbb{R})_L \times \mathbb{R}_{null}$, and these spacetimes are also known as null warped $AdS_3$. In general, they are not solutions to pure gravity alone, but can be solutions of e.g. three-dimensional gravity coupled to a massive vector field, where

\be
ds^2 = - b^2 r^2 dt^2 + \frac{dr^2}{4 r^2} + 2 r dt dy \;, \;\;\;\;\; A = b\, r dt \label{schr}
\ee
What is interesting about these backgrounds is that, even though they are not asymptotically locally $AdS_3$, they are reasonably well understood from a holographic point of view \cite{nr}.
When $b =0$, the geometry becomes $AdS_3$, so the dual description is in terms of a two-dimensional conformal field theory.
When
$b \neq 0$ but is infinitesimal - such that we can neglect the $b^2 r^2 dt^2$ term in the metric - one can use the usual 
AdS/CFT dictionary to interpret the massive vector field as a source for an irrelevant operator of dimension $(1,2)$. 
To next order in $b$ the metric backreacts but, as it has been argued in \cite{baltbob} and will be further argued in the next
 subsection, this should \emph{not} be interpreted as a source for an additional $(1,3)$ operator, but rather as a nonlinear
correction to the usual AdS/CFT dictionary. The gravity solution is in fact exact to second order in perturbation theory \cite{kraus},
 so the parameter $b$
can  take any finite value. The resulting background  \eqref{schr} has a new scaling symmetry - the so-called nonrelativistic scale
 invariance or Schr\"{o}dinger invariance - with respect to which the deforming operator in \emph{exactly marginal} \cite{nr}.
 Thus, as soon as one turns on $b$, one finds himself at a new conformal fixed point, this time with ``nonrelativistic''
scaling symmetry.

In addition, it has been argued in \cite{kraus} that for CFTs that admit a weakly-coupled dual supergravity description, the deformation by the $(1,2)$ operator is \emph{exact}, in the sense that no additional (single-trace) exactly marginal operators with respect to nonrelativistic scaling are turned on\footnote{As we will discuss in section \ref{exact}, there are subtleties related to the possible appearance of multitrace operators that one has to address.}. Consequently, all classical gravity computations in the $3d$ Schr\"{o}dinger backgrounds should be reproducible by conformal perturbation theory in the original CFT with the following deformation 

\be
S_{CFT} \r S_{CFT} + b \sqrt{N} \int dt dy \, \O_{(1,2)}
\ee
which is irrelevant from the point of view of the usual two-dimensional conformal group, but exactly marginal with respect to the nonrelativistic one. It would be very interesting to check this prediction in detail, a program which has already been started in \cite{nr}.

\bigskip

For the rest of this subsection we would like to show that in a certain $T \r 0$ limit of the backgrounds \eqref{bck}, we
recover precisely the three-dimensional Schr\"{o}dinger spacetimes \eqref{schr}. This limit should roughly correspond 
to a zero-temperature limit in the dual field theory. The parameter $T$ does 
not appear explicitly in the metric and gauge fields, but is implicit in the identifications of the coordinate $y$

\be
y \sim y + 2 \pi T \label{idyT}
\ee
When taking $T \r 0$, we must specify which quantities we keep fixed. In the first 
subsection we have argued at length that the natural boundary coordinate is not $y$,
but rather $\ty$, with identification

\be
\ty \sim \ty + 2 \pi \label{nq}
\ee
In order to keep the boundary metric finite we need to also rescale\footnote{ Note that instead of rescaling $t$ we 
could have equally well rescaled $r$ as $r = \tilde{r}/T$. The interpretation of this alternative rescaling is discussed 
in section \ref{mxlim}.} e.g. $t$ by defining $\tt = T\, t$. Finally, 
in order for the remaining metric coefficients to not blow up as $T \r 0$, we need to scale the deformation parameters as
 
\be
\e_B =   \l_B  T + \O(T^2) \;, \;\;\;\;\; \e_g =  \l_g T + \O(T^2) \label{deflam}
\ee
and fix $\l_{B,g}$ in this limit. As a result, 

\be
 \g = 1 +  (\l_B \l_g - \l_B^2) T^2 + \O(T^3)
\ee
Next, taking $T \r 0$  we obtain precisely the three-dimensional Schr\"{o}dinger spacetimes \eqref{schr} times a three-sphere

\be
ds^2 =\frac{\ell^2}{4} \left[  (\l_B \l_g-\l_B^2 - \l_g^2  ) r^2 d \tt^2 + \frac{dr^2}{r^2} + 2 r d\tt d\ty + d\th^2
 + \sin^2 \th d\phi^2 + (d\psi + \cos \th d \phi + \l_g r d\tt)^2 \right]\non 
\ee

\be
\hat F_3 + \hat H_3 = \hat F_3^{+(0)} + d B_2\;, \;\;\;\;\; B_2 = \frac{\ell^2 \l_B}{4} r d\tt \wedge \s_3 \;, \;\;\;\; e^\Phi =1 \label{schrduals}
\ee
with a null quotient given by \eqref{nq} acting on them. The dipole backgrounds have $\l_g=0$, whereas the self-dual backgrounds have  $\l_g = 2 \l_B$. These backgrounds (without the quotient) and their supersymmetry properties have been already partly analysed in \cite{kraus, baltbob} in terms of deformations of $AdS_3 \times S^3$ and consequently of the dual CFT.  In \cite{kraus}, the dipole type deformations were called diagonal, while the self-dual ones were called mixed.

Given the argument that Schr\"{o}dinger backgrounds correspond to exact deformations of the dual CFT by the operators \eqref{nso} and \eqref{sso}, both of which are supersymmetric, consistency requires that we find the same number of supersymmetries in spacetime. More precisely,  the $SL(2,\mathbb{R})_L \times SU(2)_R$ invariant Schr\"{o}dinger backgrounds should preserve four right-moving Poincar\'e supersymmetries, whereas the $SL(2,\mathbb{R})_L \times SU(2)_L$ invariant backgrounds should preserve eight superconformal left-moving supersymmetries. This is precisely what we find from the Killing spinor analysis, presented in appendix \ref{killsp}.

\subsection{Exactness of the perturbation \label{exact}}

The fact that $3d$ Schr\"{o}dinger spacetimes appear in the $T \r 0$ limit of the spacetimes we study is very interesting, as it indicates that in this limit our backgrounds are described by nonrelativistic CFTs which, at least at large $N$ and strong coupling, are exact deformations of relativistic ones. The parameter $T$ that appears in the quotient should be, roughly-speaking, the right-moving temperature in the dual field theory. If this identification is correct, then it is natural to conjecture that our backgrounds are dual to finite right-moving temperature states in the nonrelativistic CFTs dual to the backgrounds \eqref{schrduals}. 

The difficulty lies in proving that the identification of the $y$ coordinate does indeed correspond  to a right-moving temperature in the dual field theory. The proof would  amount to having a full understanding of the holographic dictionary in warped $AdS_3$, a program which has been started in \cite{nr} but is still far from completion. In lack thereof, we can use the analogy with $AdS_3$, where the fact that quotients correspond to a temperature is well-established \cite{strexcl}. Also, enticing arguments have been given that quotients of spacelike warped $AdS_3$ solutions of topologically massive gravity should correspond to a temperature in the dual field theory \cite{stromgirls}. For the special case of the dipole backgrounds, the situation is in fact better, as we can use the TsT transformation to argue that the quotient does indeed correspond to a right-moving temperature (see section \ref{dip} for details). It  would nevertheless be very interesting to show that this is the correct interpretation  also in the general case.

Even in absence of a proof, we can use the isometries of our backgrounds and the  resemblance between their asymptotic structure and that of Schr\"{o}dinger spacetimes to  directly argue that they correspond to exact deformations of the D1-D5 CFT at finite right-moving temperature.  More precisely, if we make use of the fact that all operators which are turned on (either in the form of sources or expectation 
values) must respect the $SL(2,\mathbb{R})_L \times U(1)_R$ symmetry realised by the background, we conclude that
\bi
 \item all operator deformations must be by marginal operators  with respect to $SL(2,\mathbb{R})_L$, i.e. whose scaling dimension is of the form $(1,n)$.
\item the only local operators that can acquire expectation values are those of weights $(0,n)$.
\ei

This, combined with the fact that in the supergravity limit only operators with spin at most two continue to have low dimension, implies that the only single-trace operator deformations can be by $(1,2)$ and $(1,3)$ operators, while only $(0,1)$ and $(0,2)$ operators can acquire expectation values (in a CFT these would be the right-moving currents and  stress tensor). We are for now neglecting the important issue of possible multitrace deformations (which \emph{can} have spin greater than two)  also being turned on. 

The $(1,2)$ operators in question are given by \eqref{nso} and \eqref{sso} respectively, and our next task is to argue
 that, as is true of Schr\"{o}dinger
backgrounds, no $(1,3)$ operator is turned on. The simplest way to see this is to use the specific form of our solutions:
if a $(1,3)$ operator were indeed turned on, it would correspond to a tensor mode deformation of the $AdS_3$ metric. Such a 
deformation is indeed present at second order in perturbation theory (in $\e_{B,g}$), and we have to decide whether
it is due to a nonlinear correction to the holographic dictionary for the massive vector mode, or to a new
massive graviton mode, which would represent a \emph{linearized} perturbation in $\e^2$. If the latter is the case, then we 
should find this linearized KK graviton mode in the tables of e.g. \cite{dkss}, as well as in the dual CFT. Because the 
dependence of the three-dimensional part of the metric in \eqref{bck} on the spherical coordinates is trivial, the quantum
 numbers of this putative graviton are 

\be
(h_L,h_R) = (1,3) \;, \;\;\;\;\;\;\; (j_L,j_R) = (0,0)
\ee
It turns out that  linearized spectrum of type IIB supergravity on K3 only contains fields with $h_R \leq j_R +2$, so we conclude that the only single-trace operator deformation is by the $(1,2)$ operator found in the previous section.\footnote{Another, more rigorous way of proving this would be along the lines of \cite{baltmulti,baltcallan}, who were studying irrelevant deformations by scalar operators. There it was found that at higher and higher order, the perturbation induces more and more divergent terms in the asymptotic field expansion; nevertheless, they all correspond to the \emph{same} source in the dual field theory. }

In the case of Schr\"{o}dinger backgrounds, exactness of the deformation in the CFT is nicely mirrored in the exactness of the gravity solution to second order in perturbation theory \cite{kraus}. The argument is based on the fact that the perturbation has definite and strictly positive scaling weight with respect to the Lorentz symmetry of the background. Consequently, at each subsequent order in perturbation theory the spin of the perturbation has to increase by one. Since no fields of spin greater than two are present in the supergravity limit, perturbation theory has to truncate at second order. The same argument can no longer be applied here, as the starting background has only $SL(2,\mathbb{R})_L \times U(1)_R$ isometry, and the perturbation does not in fact break any extra isometries\footnote{
A possibly amusing observation is that both the dipole and the self-dual backgrounds can be thought of as being exact to second order in perturbation theory (in gravity), provided one rescales the metric by an appropriate (fudge) factor. For the dipole backgrounds, this factor is the power of the dilaton which yields the string frame metric, and one obtains that in string frame the perturbation is exact to second order in 

\be
\e \equiv \frac{\g \, \e_B}{h}= \frac{\e_B}{\sqrt{1+ \e_B^2}} \;, \;\;\; \mbox{since} \;\;\; \g = \frac{1}{1+ \e_B^2} = 1 - \e^2
\ee
For the self-dual background, the fudge factor is $\coth 2 \d$, and the expansion parameter is again the ratio of the change in $F_3$ with respect to $F_3^{(0)}$ in \eqref{bck}

\be
\e \equiv \frac{\g \, \e_B}{h} = \frac{1}{\cosh{2 \d}} \;, \;\;\;\;\;\; \g = 1 + \frac{1}{\cosh^2 2 \d} = 1 + \e^2 \;, \;\;\;\;\;\; \g\, \e_g = 
\frac{2}{\cosh{2\d}} = 2\, \e
\ee}. Nevertheless, given that the deforming operator is still null on the boundary, most of the analysis done in \cite{kraus} for Schr\"{o}dinger backgrounds  should still apply in our case. 

Finally, let us discuss the operator expectation values. An expectation value for a $(0,1)$ operator would require the existence of a massless gauge field in the bulk solution. The analysis of section \ref{sugran} shows that no such field is turned on in our backgrounds, so we conclude there is no expectation value of a $(0,1)$ operator present. On the other hand, it is quite clear that we  should have an expectation value for a $(0,2)$ operator turned on, since the $AdS_3$ background  that we have started from had nonzero right-moving energy density corresponding to a thermal state of right-moving temperature $T_R = \frac{T}{\pi \ell R}$.
\bigskip
\bigskip

\noindent To summarize, the proposal of this section is that 

\bigskip

\parbox[c]{13 cm}{\emph{The backgrounds \eqref{bck} should be thought of as exact deformations of the DLCQ of the D1-D5 CFT by a $(1,2)$ operator where, in addition, a finite right-moving temperature $T_R$ is turned on.}}

\bigskip

\noindent Several clarifications should be made with regard to the above statement. The first issue that we have to address is what we mean by an \emph{operator deformation of the DLCQ} of the D1-D5 CFT. Despite their ubiquitousness in describing extremal black hole horizons, DLCQs of two-dimensional CFTs are not well understood. In particular, it is unclear what their spectrum of operators is, and what we mean by deforming by an operator of left/right-moving dimension $(1,2)$ a theory that is effectively one-dimensional.

A much better defined approach is to first deform the CFT$_2$ at finite right-moving temperature $T_R$ by a single $(1,2)$ operator, and only then take the DLCQ limit. Then we expect that correlators in the original CFT which have the form

\be
\left\langle T_R, 0_L | \Pi_i \O_i (x_i) \, e^{-S_{CFT} + b \sqrt{N} \int \O_{(1,2)}} |0_L, T_R \right\rangle \label{pres}
\ee
survive the DLCQ limit and agree with the correlators that we would have obtained by first taking the DLCQ limit and then deforming. Of course, we would very much like to be able to check - provided we understand the spectrum and how to deform the DLCQ - that the two procedures yield the same result, but for the time being we will use \eqref{pres} as our definition of correlators computed in the deformed DLCQ. The assumption that the DLCQ limit commutes with the deformation is supported by our concrete example in section \ref{lightdip}, where we show that this indeed happens for the gravitational dual of the lightlike dipole theory.

Another important question is whether the coefficient $b$ of the deformation depends on the temperature. 
Since finite-temperature holography in Schr\"{o}dinger backgrounds is very mysterious, at the time being we cannot employ it to answer this question. Nevertheless, for the dipole backgrounds we can show, using the TsT transformation, that the value of $b$ is independent of the temperature (see section \ref{spdipsec}). For the self-dual background, it is not known whether the coefficient of the deformation depends on the temperature, but we hope that in the future holography will provide an answer.

Finally, the most important assertion that we are making is that of \emph{exactness} of the operator deformation. Here we are employing the term ``exact'' with two distinct meanings

\bi
\item the $(1,2)$ operator is \emph{exactly marginal} with respect to the nonrelativistic scaling symmetry
\item there are no additional (exactly marginal) operators turned on
\ei
The first statement was proven in \cite{nr}, up to certain subtleties pointed out in \cite{kraus}. More
precisely, \cite{nr} showed that the nonrelativistic scaling dimension of the deforming operator is unchanged at arbitrary order in conformal perturbation theory. The subtleties are concerned with the possibility that the deformation generate a nontrivial $\b$-function for   higher-spin operators which are marginal with respect to the Schr\"{o}dinger symmetry. Given that all such operators have dimension  $(1,n)$, there is a relatively small number of candidate  operators that one can construct with these quantum numbers, especially after taking into account R-symmetry and $SO(4)_{outer}$ charge conservation. In particular, as shown earlier in this section, there is no candidate single-trace operator that has the correct quantum numbers.

The question of exact marginality is related to the types of singularities that appear when two deforming operators are brought close together and can be answered from the CFT side alone. On the other hand, the issue of exactness of the deformation (in the second sense) requires a detailed understanding of the holographic dictionary at nonlinear order.  That no additional \emph{single-trace} operators are turned on in the supergravity limit has been argued to be true for Schr\"{o}dinger spacetimes in \cite{kraus, baltbob}, and at finite temperature in the present section.  Nevertheless, we have neglected so far  the possibility that \emph{multitrace} operators\footnote{Since we are working at large N, there is a clear distinction between single-trace and multitrace operators at any point in the moduli space, consisting in whether correlation function of the given operator factorize or not \cite{shkyr}.  } may appear in the deformation. Since there is no upper bound on the spin of multitrace operators, a priori an arbitrary number of them could be present.  Also, the holographic dictionary for multitrace operators is rather subtle \cite{Witten:2001ua, Berkooz:2002ug, Aharony:2001pa, Aharony:2001dp, Mueck:2002gm, Sever:2002fk, Elitzur:2005kz, Papadimitriou:2007sj, baltcallan, baltmulti}, so their presence may be difficult to detect.

The candidate operators that could in principle spoil exact marginality are the same as those that could spoil exactness of the deformation. The only difference is that the former would appear in the action  with coefficients which are logarithmically divergent in the cutoff, whereas the latter would appear with constant coefficients. The vanishing of the coefficients of the former can be established by computing certain correlation functions, whereas for the latter one needs to compare correlation functions in supergravity with the ones in the field theory order by order in perturbation theory.

Naively, one may think that multitrace contributions can be neglected in the infinite $N$ limit in which we are working, since the coefficient with which they would appear in the action is suppressed by powers of $N$ with respect to the coefficient of the single-trace deformation. Nevertheless, it can be easily shown that the contribution of multitrace operators to certain extremal three-point correlators is of the same order as that of single-trace operators, and similar facts hold for their contribution to non-extremal higher-point functions. Consequently, the possible presence of multitrace operators in the deformation cannot be ignored. 

Let us make a few remarks about the form of the possibly ``dangerous'' multitrace operators. In the case of the $SL(2,\mathbb{R})_L \times SU(2)_R$ invariant perturbation, the deforming operator is of the form 
$\O_{(1,2)}^{(1,0)}$, where the lower indices indicate the conformal dimensions and the upper ones the R-charges. The only marginal operators that can be generated are of the form

\be
\O_{(1,n)}^{(0,0)} \;\;\;\;\;\; \mbox{or}\;\;\; \;\;\; \O_{(1,n)}^{(1,0)}\;,\;\;\;\; \;n \in \mathbb{Z}, \; n \geq 3 \label{posslr}
\ee 
based on R-charge conservation and the BPS bound. Given that all single-trace operators in the D1-D5 CFT have integer or half-integer holomorphic dimension, there are relatively few ways of constructing an operator of left-moving dimension one. Moreover, given that for a single trace chiral primary

\be
|h_L - h_R | \in \{0,1\}
\ee
it is clear that most of the single-trace operators that would be making up the multitrace have dimension $(0,1)$, i.e. they are the upper component of the right-moving SU(2) R-current, $\tilde{J}^{++}$. Consequently, computing correlation functions of the operators \eqref{posslr} should be quite straightforward, and in the end only a relatively small number of nontrivial correlators of chiral primaries will be needed for the proof of exact marginality. 

Very similar considerations apply to the $SL(2,\mathbb{R})_L \times SU(2)_L$ invariant deformation. Given that the deforming operator is now of the form $\O_{(1,2)}^{(0,1)}$, the candidate marginal operators are

\be
\O_{(1,n)}^{(0,i)} \;, \;\;\;\;\;\; i \leq n-1
\ee
As before, the fact that $h_L=1$ implies that most of the single-trace operators entering the multitrace one will be $\tilde{J}^{++}$, so correlation functions of the above operators should be again relatively easy to compute. We hope to return elsewhere to the full proof of exact marginality of the $(1,2)$ deformation of the D1-D5 CFT, as well as to the important issue whether all correlators in the deformed theory are well defined.

\bigskip

Provided that the issues mentioned above can be resolved, the fact that the backgrounds \eqref{bck} can be thought of as deformations of the DLCQ of the D1-D5 CFT by a single $(1,2)$ operator should  provide us with a  concrete framework for performing computations in the dual field theory by using conformal perturbation theory. If the latter can be well-defined, our proposal amounts to a \emph{definition} of the  field theory dual to the backgrounds \eqref{bck}, whose  properties can be studied independently of the bulk side. Computations in this field theory should reproduce the ones in classical gravity, including a field-theory derivation of the asymptotic symmetry group \eqref{asgg} and the applicability of Cardy's formula \eqref{cardy}. It would be very interesting to check whether our so-defined theories constitute an example of the $SL(2,\mathbb{R})\times U(1)$ invariant theories considered in \cite{stromhof}, which were proven to admit either two Virasoro symmetries or a Virasoro symmetry and a Ka\v{c}-Moody one. Note nevertheless that the theories defined via \eqref{defnrcft} are not guaranteed to be local, so it is not a priori clear that they should satisfy the assumptions of \cite{stromhof}.

\section{Computations in the D1-D5 CFT \label{freeorb}}

The deforming operators found in section \ref{qno}, being descendants of chiral primaries, have a well-defined meaning at any point moduli space. Nevertheless, sometimes it is useful to have explicit expressions for them. Thus, in  this section, we map the operators that we have found to specific linear combinations of twist operators at the free orbifold point of the D1-D5 CFT.

\subsection{A few facts about chiral primaries}

The CFT$_2$ dual to the $AdS_3 \times S^3 \times K3$ background of type IIB string theory is the D1-D5 CFT, an
$\N=(4,4)$ superconformal field theory whose moduli space is locally of the form \cite{dijkinst}

\be
\frac{SO(4,21)}{SO(4) \times SO(21)}
\ee
There exists a special point in this moduli space where the D1-D5 CFT has a simple description in terms of a free superconformal sigma model whose target space is the symmetric product of $N=Q_1 Q_5$ copies of K3 \cite{vafagas,microorigin}.  Applying symmetric orbifold techniques to a free field realisation, the chiral primary operators can be easily constructed \cite{Vafa:1994tf, strexcl} and their correlation functions explicitly computed \cite{Jevicki:1998bm, Lunin:2000yv, Lunin:2001pw, pakmanext, pakmandiag}. Even though the free orbifold point may be a singular point in the moduli space of the D1-D5 CFT \cite{Seiberg:1999xz}, it has been shown that the three-point correlation functions  of chiral primaries computed at this point perfectly match those computed in the worldsheet WZW model CFT \cite{atish,gargamel}, as well as those computed in supergravity \cite{marika}. We are clearly dealing with an $\N=4$ nonrenormalization theorem for the above correlators\footnote{This nonrenormalization theorem has been proven for all points in the moduli space and at finite $N$ in the special case of extremal correlators \cite{kyriakos}. }.

This nonrenormalization theorem can be used to map the chiral primary operators e.g. in the region of the moduli space which has a weakly-coupled supergravity description to e.g. those at the free orbifold point. As pointed out in \cite{kyriakos} this map is not unique, due to operator mixing under flow in the moduli space. As shown in the same paper, the chiral primaries of a given conformal dimension in the  $\N=(4,4)$ SCFT are sections of vector bundles over the moduli space,  which have constant but nonzero curvature. Due to the curvature, the identification of the operators between two points in the moduli space depends on the path taken to connect them. The connection on the vector bundle of chiral primaries has $SO(21)$ holonomy. Consequently, the ambiguity in the identification will exist for all chiral primary operators that transform in nontrivial representations of $SO(21)$, whereas operators which are  $SO(21)$ singlets will be unambiguously identified.  

The map between supergravity and free orbifold CFT chiral primaries has been constructed in \cite{marika}. Before  proceeding to describing it, let us remind the reader how chiral primary operators in the free orbifold CFT are  constructed. The free orbifold CFT contains $4N$ free bosonic fields  $X^{m,\bar m}_A$  their left-moving fermionic superpartners $\psi^{m \a}_A$ and right-moving fermionic superpartners $\psi^{\bar m \dot \a}$. Here $m$ is a holomorphic index on $K3$ taking values $m = \{1,2 \}$, $\bar m$ is an antiholomorophic one, $A = 1, \ldots , N$ labels the copies of K3, $\a$ is an $SU(2)_L$ R-symmetry index and $\dot \a$ is an $SU(2)_R$ one.

The chiral primary operators are constructed from the fields $X_A$, $\psi_A$ and $\tilde{\psi}_A$ with either twisted or untwisted boundary conditions. In the untwisted sector, since only $\psi_A, \tilde \psi_A$ carry R-charge, the chiral primaries are in one to one correspondence with the cohomology classes of K3 

\be
\O_1^{(1,1)\,S} = \frac{1}{\sqrt{N}} \sum_A \om^{S}_{m \bar m} \psi_A^{m +} \tilde{\psi}_A^{\bar m + }\;, \;\;\;\;\; 
\O_1^{(2,2)} = \frac{1}{\sqrt{N}}
\sum_A \om^{(2,2)}_{m n \bar m \bar n} \, \psi_A^{m+} \psi_A^{n+} \tilde{\psi}_A^{\bar m +} \tilde{\psi}_A^{\bar n +} \non 
\ee
\be
\O_1^{(2,0)} = \frac{1}{\sqrt{N}} \sum_A \om^{(2,0)}_{m n} \psi_A^{m +} \psi_A^{n + } = J^{++} \;, \;\;\;\;\;
\O_1^{(0,2)} = \frac{1}{\sqrt{N}} \sum_A \om^{(0,2)}_{\bar m \bar n} \tilde{\psi}_A^{\bar m +} \tilde{\psi}_A^{\bar n + } = \tilde{J}^{++} \label{11cp}
\ee
Here the index $S = 1, \ldots, 20$ runs over the $(1,1)$ harmonic forms on K3. The operators constructed
from the $(2,0)$ and $(0,2)$ forms on K3 are nothing but the upper component of the left/right-moving $SU(2)$ R-symmetry currents, as indicated. Thus, from the untwisted sector we get $20$ $(\half,\half)$ chiral primaries, one $(1,1)$ one, and in addition two currents, one $(1,0)$ and one $(0,1)$. The subscript `$1$' on the operators indicates the fact we are in the untwisted sector, whereas the upper scripts indicate the number of fermionic creation operators we have used on each side.

Nevertheless, most of the chiral primaries in the CFT are constructed from fields with twisted boundary conditions. 
The twist fields of the SCFT are labeled by the conjugacy classes of $S(N)$, namely the cyclic groups of 
various lengths $Z_n$, with $n \leq N$. A $Z_n$ twist yields a chiral primary $\Sigma^{(\frac{n}{2}, 
\frac{n}{2})}$ of dimension

\be
\mbox{dim} \left( \Sigma^{(\frac{n}{2}, \frac{n}{2})} \right) = \left( \frac{n}{2}, \frac{n}{2}\right)
\ee
In addition, we can combine these twist operators with the fermionic creation operators as in the untwisted sector. The general structure of an operator is then $\O_n^{(a,\bar{a})}$, where $n$ indicates the twist and $0 \leq a,\bar a \leq 2$ indicates the number of fermionic creation operator insertions.
For $a=\bar a =1$ we have an additional index, $S$, which labels the $20$ $(1,1)$ cohomology classes of $K3$.

\subsection{Identification of the operators}

The conclusion of the supergravity analysis of  section \ref{qno} was that the $SL(2,\mathbb{R})_L \times SU(2)_R$ invariant deformations correspond to operators of the form  

\be
\O_{(1,2)}^{{}^{LR}}=J^{--}\tilde{G}^{--} \tilde{G}^{-+} \O_{(1,1)}^\chi
\ee
where $\O_{(1,1)}^\chi$ is a dimension $(1,1)$ chiral primary which is an $SO(21)$ singlet in case of the self-dual deformation and a  $SO(21)$ vector for the case of the nonsupersymmetric dipole backgrounds. Note that for simplicity reasons we have suppressed the subscripts $-\half $ on the supercharges, and we will continue to do so throughout this section. 

For the superconformal  dipole deformations, the dual operators take the form

\be
\O_{(1,2)}^{{}^{LL}} =\tilde J^{--} G^{-a}\tilde{G}^{-b} \O_{(\half, \frac{3}{2})}^\chi
\ee
where $\O_{(\half, \frac{3}{2})}^\chi$ is the single chiral primary of this dimension in the CFT. What we would like
to do now is to give explicit expressions for the 22 chiral primary operators $\O_{(1,1)}^\chi$ and the operator $\O_{(1/2, 3/2)}^\chi $ in the free orbifold CFT.

The list of  $n+1=22$ chiral primary operators of dimension $(1,1)$ in the free orbifold CFT is, schematically

\be
\O_3^{(0,0)} = \Sigma^{(1,1)} \;, \;\;\;\;\; \O_2^{(1,1) \, S} = \Sigma^{(\frac{1}{2}, \frac{1}{2})} \,
 \om^S_{(1,1)} \, \psi^+ \tilde{\psi}^{ +}\;, \;\;\;\;\; \O_1^{(2,2)} = 
\sum_A \om_{(2,2)} \,\psi^+ \psi^+ \tilde{\psi}^+ \tilde{\psi}^+
\label{onecone}
\ee
where appropriate summations over $Z_2$ cyclic permutations in $S_N$ and over copies of $K3$ are assumed\footnote{The issue of which is the appropriate summation to perform for the second set of operators is subtle and has been resolved in \cite{atish}. The exact expression for the third operator is given in
\eqref{11cp}. }. The $(\half, \frac{3}{2})$ operator relevant for the superconformal dipole deformations is schematically given by

\be
\O^\chi_{\left(\half, \frac{3}{2} \right)}= \O_2^{(0,2)} = \Sigma^{(\half,\half)}  \om_{(0,2)}\, \tilde{\psi}^{+} 
\tilde{\psi}^{ +}
\ee
The degeneracies of these operators agree with those found from the supergravity analysis. Namely, in the $SL(2,\mathbb{R})_L \times SU(2)_R$ invariant case we have $22$ deformations induced by the massive KK modes of the fields

\be
F_3^+\;, H_3^- , \; F_3^- \;,  F_3^{\hat{S} - } \;, \;\;\;\;\; \hat{S} = 1 , \ldots , 19 
\ee
where the index $\hat{S}$ labels the $19$ anti-self-dual $(1,1)$ forms in $K3$. We label the $(1,1)$
holographic chiral primary operators whose descendants are dual to the above deformations by 

\be
\O_{F^+_3}, \O_{H^-_3} \;\; \mbox{and} \;\; \O^S_{F^-_3}
\ee
Here the index $S = 1, \ldots, 20$ comprises both $F_3^{\hat S}$ and $F_3^-$, as the two transform into each other under the $SO(20)$ subgroup of the $SO(4,20)$ T-duality group of type IIB on K3. The descendant of $\O_{F_3^+}$ is responsible for the self-dual deformation and corresponds to the $SO(21)$ singlet, whereas the supersymmetric descendants of all the remaining operators represent dipole deformations, transforming in the vector representation of $SO(21)$. 

The question now  is how to map the above supergravity operators to the free orbifold CFT operators \eqref{onecone}. One way of identifying the operators on the two sides  is to use the nonrenormalization of the nonextremal three-point correlation functions of chiral primaries\footnote{ This identification 
of single-trace operators on the two sides is only valid at large $N$. The map gets corrected by multitrace operators at subleading order in $N$, but the corrections only affect the extremal three-point correlators, which is the reason why \eqref{opmap} has been established using the matching of the  nonextremal ones.
} \cite{marika}. The proposed map is

\be
\O_{F^+_3} \r \frac{\sqrt{3}}{2} \,\O_1^{(2,2)} + \half\, \O_3^{(0,0)} \;, 
\;\;\;\;\; \O_{H^-_3} \r \frac{\sqrt{3}}{2} \,\O_3^{(0,0)} - \half \, \O_1^{(2,2)} \non
\ee

\be
\O^S_{F_3^-} \r \O_2^{(1,1)\, S} \label{opmap}
\ee
  It is probably interesting to note that the $SL(2,\mathbb{R})_L \times SU(2)_R$ invariant self-dual deformation $\O_{F^+_3}$, which we will soon show is the one relevant for understanding Kerr/CFT, is a linear combination of the same operators as the $SL(2,\mathbb{R})_L \times SU(2)_R$ invariant dipole deformation. Since dipole deformations of the D1-D5 CFT can be in principle understood starting from the star product deformation of the D1-D5 gauge theory Lagrangian  and flowing to the IR, one may hope that a similarly simple interpretation of Kerr/CFT for this system can be found. 

As far as the four $SL(2,\mathbb{R})_L \times SU(2)_L$ invariant  dipole deformations are concerned, we are supposed to map the four operators $\tilde{J}^{--} G^{-a} \tilde{G}^{-b} \O_2^{(0,2)}$ onto the KK modes of the four supergravity fields

\be
H^+_3, F_3^{A+}
\ee 
This map is fixed by supersymmetry. Namely, the operators $G^{-a} \tilde{G}^{-b} \O_2^{(0,2)}$ 
transform in the $(2,2)$ representation of $SU(2)_L^{outer} \times SU(2)_R^{outer}  \cong SO(4)_{outer}$ of the supersymmetry algebra. The outer isomorphism group in the CFT is identified with the $SO(4)$ subgroup of the $SO(5,21)$ symmetry of $6d$ supergravity which is not broken by the background expectation value of $F_3^{+(0)}$. 
 Under it, the four fields $F_3^{A+}, H_3^+$ transform as a vector. 

As it was mentioned at the beginning of this section, the map between chiral primary operators which transform in nontrivial representations of $SO(21)$ and are at different points in the moduli space  should be inherently ambiguous due to the curvature of the moduli space. Therefore, the map \eqref{opmap} does not in fact make sense without a specification of the path connecting the two points in the moduli space. While the identification of $\O_{F^+_3}$ is unambiguous due to the fact that it is an $SO(21)$ singlet - which is good news from the point of view of Kerr/CFT - the map for the dipole deformations should in general be modified to

\be
\left(\begin{array}{c} \O_{H^-_3}  \\ \O^S_{F_3^-} \end{array}\right) \longleftrightarrow  \; \mathcal{M} 
\left(\begin{array}{c} \half( \sqrt{3}\, \O_3^{(0,0)} -  \O_1^{(2,2)}) \\  \O_2^{(1,1)\, S} \end{array}\right)\;,\;\;\;\;\;\;\;\;
\mathcal{M} \in SO(21)
\ee
The $SO(21)$ matrix $\mathcal{M}$ depends on the path through the moduli space taken from the supergravity point to the free  orbifold point. Since the connection on the vector bundle of chiral primaries over the moduli space has full $SO(21)$ holonomy, we expect to always be able to find a path such that $\mathcal{M}$ is the identity matrix. In fact, given the naturalness of identifying the spacetime indices $S$  which transform under the $SO(20)$ subgroup of the T-duality group with the $SO(20)$ which acts on  the indices of the $(1,1)$ forms on K3 inside the operators \eqref{onecone}, one may wonder whether the path for which $\mathcal{M} = \mathbb{I}$ may have a very simple interpretation. For example, it could correspond to a geodesic, or to an orbit of an isometry of the moduli space\footnote{We thank K. Papadodimas for these suggestions.}. It would be very interesting to further study these issues, but they go beyond the scope of the present article.  

The same comments apply to the mapping of the $SL(2,\mathbb{R})_L \times SU(2)_L$ invariant deformation operators. In this case, the chiral primary whose descendants we are interested in is unambiguously identified between supergravity and the CFT, but the map for the full operator \eqref{sso} does depend on the path through moduli space that we take because the supersymmetry generators themselves have nontrivial $SO(4)_{outer}$ holonomy.  Note that the full $SL(2,\mathbb{R})_L \times SU(2)_R$ invariant deformation \eqref{nso} is a singlet under $SO(4)_{outer}$, so the only ambiguity in the identification is the $SO(21)$ discussed above.

Another interesting observation, which is at the core of \cite{marika}, is the fact that the map \eqref{opmap} receives subleading corrections in $1/N$ from multitrace operators. These corrections do in fact affect certain extremal correlators of chiral primary operators, which naively do not match between the supergravity and the free orbifold point. These correlators can consequently be used to infer the corrections from multitrace operators to the map \eqref{opmap}. The computations of \cite{marika} show that the $(1,1)$ chiral primary operators that enter \eqref{opmap} are corrected as

\be
\O^\chi_{(1,1)} \r \O^\chi_{(1,1)} + \frac{\a}{\sqrt{N}}\, \S^{(\half,\half)} \S^{(\half,\half)} \label{mtcorr}
\ee
where $\S^{\left(\half,\half\right)}$ is the $Z_2$ twist operator in the free orbifold CFT and the numerical coefficient $\a$ depends on whether the chiral primary in question transforms as a singlet or vector of $SO(21)$. One can in principle check whether the above correction is consistent with the $SO(21)$ transformation properties of the operators. Namely, the map for $(1,1)$ chiral primaries can receive subleading  corrections in $1/N$ from  double trace operators of lower dimension. The double trace operator can be constructed from $(\half,\half)$ chiral primaries, which are $21$ in number and transform as a vector under $SO(21)$. Consequently, the double trace transforms in the tensor product of two such vector representations, namely

\be
21 \otimes 21 = 210 \oplus 21 \oplus 1
\ee
Only the $21$ on the right-hand side can correct the $(1,1)$ chiral primaries which transform in the vector representation, wheres only the $1$ can correct the singlet $(1,1)$ operator. Note that the $SO(21)$ singlet chiral primary can also in principle receive contributions from the multitrace operator 
$J_L^{++} J_R^{++}$, but \eqref{mtcorr} shows that it does not.

\section{Relationship to the D1-D5-p black hole} \label{d1d5rev}

In this section we review how the self-dual deformation is precisely the one that arises in the near-horizon limit of the D1-D5-p black hole/string with equal D1 and D5 charges. We also explain the slight difference between the proposal of \cite{microkerr} and the computations performed in the present article.

\subsection{The near-horizon geometry }

Consider a D1-D5-$p$ black hole in type IIB supergravity compactified on $K3 \times S^1$ or $T^5$.
We set the right-moving angular momentum $J_R$ to zero, but let $J_L \neq 0$. 
The charges of the extremal nonsupersymmetric
black hole can be parametrised as\footnote{We use the conventions of 
\cite{kerrscatt}. In order to switch from $J_L=0$ to $J_R=0$ we simply replace $\d_p \r - \d_p$.}

\be
Q_1 = 2 a^2 \sinh 2 \d_1\;,\;\;\;\;\;Q_5 = 2 a^2 \sinh 2 \d_5\;,\;\;\;\;\;Q_p = 2 a^2 \sinh 2 \d_p\;,\;\;\;\;\;
\ee
while the mass and angular momentum are

\be
M = 2 a^2 (\cosh 2 \d_1 + \cosh 2 \d_5 + \cosh 2 \d_p) \;, \;\;\;\;\; J_L = 4 a^3 (c_1 c_5 c_p + s_1 s_5 s_p)
\ee 
We have used the standard shorthand $c_i = \cosh \d_i$ and $s_i = \sinh \d_i$. The  entropy of the black hole is given by

\be
S= 2 \pi \sqrt{J_L^2 - Q_1 Q_5 Q_p}
\ee
The near-horizon limit of this black hole is analysed in \cite{emparan,kerrscatt}. It takes the form 

\bea
ds_6^2 & = & \frac{K_0}{4} \left(-r^2 dt^2 + \frac{dr^2}{r^2} + \g (dy+rdt)^2 + \g (d\psi + \cos\th d\phi)^2  + \right.\non \\
&& \hspace{3 cm}  2 \a (dy+rdt)(d\psi + \cos\th d \phi) + d\th^2 + \sin^2\th d\phi^2 \biggr) \label{nhgk}
\eea
This expression is the same as that in \cite{microkerr}, except that the formulae for $K_0, \a, \g$ are generalized to 

\be
\g = 1+\frac{1}{ \cosh 2 \d_1 \cosh 2 \d_5}\;, \;\;\;\;\;\a = \frac{1}{\cosh 2 \d_1} + \frac{1}{\cosh 2 \d_5}
\ee

\be
K_0 = 2 a^2 \sqrt{\cosh 2 \d_1 \cosh 2 \d_5} \label{K0}
\ee
Using \eqref{gamh}, the RR three-form field strength reads 

\bea
F_3 & = & \frac{K_0 \tanh 2 \d_5}{4} \left(  \sin \th d\th \wedge d\phi \wedge d\psi + dr\wedge dt \wedge dy  \right. +
\hspace{5 cm}\non \\
&&  \hspace{2 cm}\left. +\; \frac{ \sin \th d\th \wedge d\phi \wedge (dy + r dt) + dr \wedge dt \wedge (d\psi + \cos \th d\phi)}{\cosh 2 \d_1}  
\right)
\eea
Note that when $Q_1=Q_5 =Q$ then $F_3$ is self-dual. Letting $\d_1 = \d_5 = \d$, the above background is \emph{locally} identical to the self-dual background \eqref{bck} - \eqref{sdcond}.

The coordinates $y$ and $\psi$ have identifications\footnote{In terms of the coordinates $\tilde{\psi}$ and $\tilde{y}$, which are identified mod $4\pi$ and respectively $2\pi$ and are  appropriate near-horizon coordinates related to the angular direction on the $S^3$ and the $6d$ $U(1)$ fibre by a singular shift in $t$, the coordinates $\psi$ and $y$ are given by 
\be
\psi = \tilde{\psi} - \frac{c_1 c_5 c_p +s_1 s_5 s_p}{2 a s_1 s_5 c_1 c_5}\, \tilde{y} \;, \;\;\;\;\; y = \frac{c_1 c_5 c_p - s_1 s_5 s_p}{2 a \, c_1 c_5 s_1 s_5} \,\tilde{y}
\ee
The coordinates $\tilde{y}$ and $\tilde{\psi}$ are defined in section 6.1.2 of \cite{kerrscatt}, with the mention that $\tilde{y},\tilde{\psi}$ in that paper correspond to our $y,\psi$ and vice-versa.  }

\be
y \sim y + 4 \pi^2 T_Q m \;, \;\;\;\;\; \psi \sim \psi + 4 \pi n - \frac{4 \pi J_L m}{Q_1 Q_5} \label{identif}
\ee
where $T_Q$ is given by

\be
T_Q = \frac{J_L T_R}{Q_1 Q_5} \;, \;\;\;\;\;\; 
T_R =\frac{c_1 c_5 c_p - s_1 s_5 s_p}{\pi (c_1 c_5 c_p + s_1 s_5 s_p)} =\frac{1}{\pi} \sqrt{1-\frac{Q_1 Q_5 Q_p}{J_L^2}}
\ee

\subsection{Holographic interpretations \label{holoint}}

 In \cite{microkerr} it has been argued that in the maximally charged limit

\be
Q_1 Q_5 Q_p \r J_L^2 \label{mxl}
\ee
the near-horizon geometry \eqref{nhgk} contains an $AdS_3$ factor, and thus it is possible to study it using the usual
rules of AdS/CFT. Indeed, this limit corresponds to

\be
\d_i \r \infty\;, \;\; a \r 0\;\; \;\;\;\; \mbox{with} \;\;\; Q_1= a^2 e^{2\d_1}, \;\; Q_5 =   a^2 e^{2\d_5}, \;\; 
Q_p = a^2 e^{2\d_p}= \frac{J_L^2}{Q_1 Q_5}
\ee
fixed. Thus, $\a \r 0$ and $\g \r 1$ in \eqref{nhgk}, yielding a locally $AdS_3 \times S^3$ geometry
of radius $\ell^2 = \sqrt{Q_1 Q_5}$. Moreover, the identification of $\psi$ in this limit

\be
\psi \sim \psi + 4 \pi n - \frac{4 \pi m}{N}\;, \;\;\;\;\;\;\;\;N \equiv \frac{Q_1 Q_5}{J_L} = \sqrt{\frac{Q_1 Q_5}{Q_p}}  
\ee
can be interpreted as a $Z_N$ quotient of the three-sphere

\be
\psi \sim \psi - \frac{4 \pi \hat m}{N} \label{idpsi}
\ee
provided that  $N \in \mathbb{Z}$ and we use instead

\be
\hat m = m - n N
\ee
The $Z_N$ quotient of the three-sphere physically  corresponds to a charge
$N$ magnetic monopole. In terms of $\hat m$ and $n$, the identification of $y$  reads

\be
y \sim y + 4 \pi^2  T_Q \, \hat m + 4 \pi^2 T_R \,  n   \;, \;\;\;\;\;\;\;\; T_Q = \frac{T_R}{N} \label{idy}
\ee
In the maximal limit \eqref{mxl}, both $T_Q$ and $T_R$ vanish. The vanishing of $T_Q$ is not worrisome, as it simply indicates that the $\hat m$ quotient acts only on $\psi$ in this limit. The remaining identification by  $4 \pi^2 T_R \,n$  yields however a singular  quotient of $AdS_3$, of the pinching orbifold type.

This $AdS_3 \times S^3/Z_N \times K3$ solution of type IIB supergravity arises as the near-horizon limit 
of a stack of D1-D5-KK branes, lying in the following configuration

\medskip

\begin{center}
\begin{tabular}{c|c|cccccc} 
&  & 6 & 7& 8 & 9 & $\psi$ & $ \mathbb{R}_{y} $  \\ \hline
$q_5$ &D5 & x & x& x & x &  & x  \\
$q_1$& D1 & & & & & & x  \\
$N \,$&TN & x&x &x &x &  &x 
\end{tabular}
\end{center}
\medskip
where $\mathbb{R}_{y} $ denotes the common string direction, $\psi$ is the Taub-Nut circle and $6,7,8,9$ denote the 
directions along K3. The number of branes of each type is

\be
q_1 = \frac{Q_1}{N}\;, \;\;\;\;\; q_5 = \frac{Q_5}{N}
\ee
and we are assuming that $N$ divides both $Q_1$ and $Q_5$. At low energies, the above brane configuration is described by a CFT known as the ``quiver projection'' of the D1-D5 CFT \cite{sugawara}, whose central charge is 

\be
c_L = c_R= 6 \, q_1 q_5 N = \frac{6 \,Q_1 Q_5}{N}=6 \sqrt{Q_1 Q_5 Q_p} = 6 J_L
\label{kerrcc}
\ee
If  $N > 1$, this CFT only has $(0,4)$ supersymmetry. Its operator content can be obtained from that of the D1-D5 CFT by a procedure similar to orbifolding described in \cite{Douglas:1996sw, sugawara}. 

It was noted in \cite{microkerr} that the central charge \eqref{kerrcc} agrees perfectly with the central extension of the Virasoro asymptotic symmetry group 
of the five-dimensional D1-D5-p black hole. Thus, in this instance of Kerr/CFT the CFT is, in the maximal limit, precisely the  DLCQ  of the quiver projection of the  D1-D5 CFT (which is related by  dualities with the DLCQ of the MSW CFT discussed in that paper). Also, in this limit 
the CFT is  in its ground state.

Going away from the maximal limit \eqref{mxl}, the $AdS_3 \times S^3/Z_N $ geometry gets deformed. The deformation consists both of a change in the local geometry - from $AdS_3 \times S^3$ to warped $AdS_3$ times a stretched sphere which is nontrivially fibered over it - and a change in the identifications. 
Infinitesimally away from the maximal limit, the deformations can be understood using the usual AdS/CFT dictionary. Given that for $Q_1 =Q_5$ the local geometry is precisely that of the self-dual backgrounds \eqref{bck}-\eqref{sdcond}, the holographic analysis of the infinitesimal change in the geometry is almost identical to that of section \ref{sugra}, and one finds a deformation by a specific $(1,2)$ operator in the quiver-projected D1-D5 CFT.  In addition, one finds that the right-movers are excited at the infinitesimal temperature $T_R$. Far away from maximality, the form of the geometry indicates that  both a deformation and a temperature are present\footnote{In the analysis of \cite{microkerr} all charges were taken to be equal, $Q_1 = Q_5 = Q_p$, and so $\d_1 = \d_5=\d_p$. Because of this, the deformation parameter, which depends on $\d_1$ and $\d_5$, and the temperature $T_R$, which depends on all three $\d_i$, could not be disentangled. Our slightly more general analysis shows that at least at the level of the analysis in \cite{microkerr}  the two deformations - operator and temperature - are in fact independent. Nevertheless, one may also want to keep other quantities - e.g. $N$ and $J_L$  - fixed, case in which the two deformations are again entangled.}, but a more precise understanding of the holographic meaning of the identifications is needed.

The holographic interpretation away from maximality depends on the way one leaves the maximal limit. In 
the three-charge case under study here the most natural way to leave maximality seems to be by lowering $Q_p$, with $J_L, Q_1,Q_5$ and thus $N$ kept fixed. In this way, the identification of $\psi$ is always given by \eqref{idpsi} with $N = Q_1 Q_5/J_L$. The term proportional to $n$ in the $y$ identification 
 \eqref{idy} should most likely still be interpreted as a temperature $T_R$ for the right-movers. Nevertheless, the holographic interpretation of the additional shift by $\hat m$ is unclear in this picture, except when $N=1$. In that case, we can simply define

\be
\hat n = n + \hat{m}
\ee
such that the identification of $y$ becomes

\be
y \sim y + 4 \pi^2 T_R \; \hat n
\ee
which is naturally interpreted as a right-moving temperature.

\bigskip

It is quite common for black holes in string theory to have several possible microscopic descriptions related by U-duality. The same happens here, as the maximal geometry also has a description in terms of the usual D1-D5 CFT with no quiver projection, but with the shift by $m$ \eqref{identif} in $\psi$ interpreted instead as a chemical potential for the left-movers. As explained  in \cite{microkerr}, in the maximal limit the quiver projection description represents the long string \cite{maldasuss} sector of D1-D5 description, as the central charges are

\be
c_{D1D5} = 6 Q_1 Q_5 \;, \;\;\;\; c_{quiv.\,proj.}= \frac{6 Q_1 Q_5}{N} = 6 J_L
\ee
Nevertheless, away from maximality we do not expect the two descriptions to be so simply related. The key difference
 lies in the interpretation of the identifications \eqref{identif}: in the D1-D5 description, the temperature for the right-movers is $T_Q$, whereas the extra shift in $\psi$ represents a chemical potential for the conjugate charge. In the quiver projection description, we interpret the entire identification \eqref{idpsi} of $\psi$ as a charge $N$ magnetic monopole, and the shift \eqref{idy} in $y$ as a temperature, plus an additional identification related to the $Z_N$ orbifold which is only present away from maximality and whose physical meaning is unclear. Both descriptions are valid, despite the different interpretations.

While the quiver projected D1-D5 description with central charge $c_{q.p.} = 6 J_L$ is more appropriate for making the connection with Kerr/CFT, for the purposes of this article we are more interested in the D1-D5 description, which is simpler from several viewpoints. First, the interpretation of the identifications away from maximality is much clearer. Second, the CFT that we are deforming in this picture has $(4,4)$ instead of just $(0,4)$ supersymmetry, and thus we gain in computational control.
Third, it is very clear in this picture that the deformation, the temperature and the chemical potential are completely independent quantities. 

The background \eqref{nhgk} with $Q_1=Q_5 =Q$ is identical to the self-dual background \eqref{bck}-\eqref{sdcond}, except for the extra shift in $\psi$ \eqref{identif}. Given that this shift simply represents a spectral flow  in the dual theory (and thus, a constant shift in the energies and currents), the analysis of section \ref{sugra} applies in its entirety for this description. Consequently, the nonsupersymmetric D1-D5-p black hole should be thought of as a thermal state in a non-relativistic CFT, which is an exact supersymmetric deformation of the D1-D5 CFT by the unique $(1,2)$ operator of the form \eqref{nso} which is an $SO(21)$ singlet. 

The asymptotic symmetry group analysis of section \ref{asg} indicates that the deformed theory has a right-moving Virasoro symmetry with central extension

\be
c = \frac{3 \pi^2 \ell^4}{G_6} 
\ee
In the parametrization used in this section, $G_6 = \frac{\pi^2}{2}$ and the central charge turns out to be

\be
c_{D1D5} = 6 Q^2 = 6 Q_1 Q_5
\ee
The above central charge, together with  the temperature $T_Q$, exactly reproduces the entropy of our original black hole by use of Cardy's formula

\be
S_{CFT} = \frac{\pi^2}{3} c_{D1D5} T_Q = 2 \pi \sqrt{J_L^2 - Q_1 Q_5 Q_p} \label{sbh}
\ee
In fact, there is a nicer and more physical way of understanding the above formula \cite{maccarrone} in the D1-D5 description. The asymptotic symmetry group analysis indicates that Cardy's formula is applicable even in the deformed CFT, and that the central charge is unchanged. Given that only right-movers are excited, the entropy is given by 

\be
S = 2 \pi \sqrt{\frac{c N_R}{6}} \;, \;\;\;\;\;\;\; c = 6\, Q_1 Q_5
\ee
where $N_R$ is the right-moving energy. Since the black hole has nonzero angular momentum $J_L$ but zero
left-moving temperature, the left-moving energy is given by

\be
N_L = \frac{6 J_L^2}{c} \;, \;\;\;\;\;\; \ni \;\; \hat N_L = N_L - \frac{6 J_L^2}{c} =0
\ee
where $\hat N_L $ is  the effective left-moving momentum to be distributed among oscillators. 
The difference between the left- and right-moving momenta is the measured charge $Q_p$

\be
Q_p = N_L - N_R =  \frac{6 J_L^2}{c}  - N_R
\ee
Solving  for $N_R$ and plugging the result into Cardy's formula, we obtain precisely \eqref{sbh}.

\subsection{A different maximal limit \label{mxlim}}

We have just shown that the self-dual background is precisely the background relevant for Kerr/CFT, barring a
 few slightly different identifications of the coordinates. In section \ref{3dsch} we have taken the zero -``temperature'' 
limit of the self-dual background and we have obtained a Schr\"{o}dinger spacetime. In the previous subsection 
we have taken the maximal limit of NHEK, which we argued is also a zero-temperature limit, but instead we have
 obtained $AdS_3\times S^3$. Clearly, we are keeping different quantities fixed in the two cases, and in this
 subsection we would like to understand the difference between the two limits.

The difference clearly lies in the fact that in the maximal limit \eqref{mxl} we have not rescaled $y,t$ before sending
$\d_i \r \infty $. Introducing the rescaled coordinates

\be
\ty = \frac{y}{2 \pi T_Q} \;, \;\;\;\ty \sim \ty + 2 \pi \;, \;\;\;\;\;\; \tt = 2 \pi T_Q \, t
\ee
and then taking $\d_i \r \infty$, we obtain a Schr\"{o}dinger background, as expected, with 

\be
\l_B =  \frac{Q \sqrt{Q_p}}{Q+2 Q_p} = \frac{Q^2 J_L}{Q^3 + 2 J_L^2} \;, \;\;\;\;\; \l_g = 2 \l_B 
\ee
where $\l_{g,B}$ have been defined in \eqref{deflam} and $Q_1=Q_5=Q$.

Note that in both zero-temperature limits ($AdS_3$ and Schr\"{o}dinger), the geometry contains a circle of vanishing 
size, so the gravity analysis is not particularly trustworthy in either case: the $AdS_3$ geometry we obtained in the 
previous section is the pinching orbifold, which has a spacelike circle of vanishing size, whereas the Schr\"{o}dinger 
geometry we uncovered has a compact null coordinate ($\ty$). There is no reason that light stringy modes should not make their appearance in either case.

Nevertheless, the gravity analysis yields rather sensible results, so we will ignore for the moment this important
issue. Then, in each case we have an operator deformation of the CFT dual to $AdS_3$ which is an irrelevant operator
 of weight $(1,2)$. Near $T_Q =0$, the coupling constants of this operator in the two
descriptions are related by

\be
\e_B = 2 \pi T_Q\, \l_B
\ee
In section \ref{sdporb}, we have argued that the most natural holographic coordinates for the DLCQ of $AdS_3$ are
 $\ty, \tt$, with $\ty \sim \ty + 2 \pi$. This implies that the true coupling constant  of the operator, at least
in the zero-temperature limit, is $\l_B$. Sending $T_Q \r 0$ with $\l_B$ fixed then automatically sends
 $\e_B\r 0$,
which explains why in the maximal limit of the previous section the operator deformation was vanishing.

Another way of understanding this relation is to reach the Schr\"{o}dinger limit not by rescaling $y$ and $t$, but rather 
by rescaling $y$ and $r$. If we define $\ty$ as above and let 

\be
\tilde{r} = 2 \pi T_Q \, r
\ee
then we obtain the same Schr\"{o}dinger geometry. Nevertheless the $T_Q \r 0$ limit with $r$ fixed now sends the holographic
RG scale $\tilde{r} \r 0$, i.e. towards the deep IR of the original theory. Since the deforming operator is irrelevant, it
vanishes in the IR, so it is natural recover the $AdS_3$ geometry dual to the CFT that we had been deforming. 

Each of the above descriptions has its advantages and disadvantages. The Schr\"{o}dinger description is more natural
from the boundary point of view and it gives us the correct normalization of the operator's coupling constant. On the 
other hand, holography is very poorly understood for this spacetime, and in particular we have no proof that $T_Q$ should be interpreted as a right-moving temperature.
The $AdS_3$ description is valid in the extreme IR of the theory, where there are relatively few states left. Nevertheless,
holography is well-understood, we do have an interpretation for $T_Q$ as the right-moving temperature, and we are entitled to
use Cardy's formula to obtain the near-maximal entropy. As mentioned above, in both pictures the gravity description is not a priori reliable, due to the presence of a vanishing-size circle in the geometry. 

Needless to say, the analysis of this section applies without change to all the backgrounds we have been studying.

\section{Interpretation of the dipole backgrounds \label{dip}}

In section \ref{exact} we have argued that the backgrounds we study are dual to exact deformations by a $(1,2)$ operator of the \emph{DLCQ} of the D1-D5 CFT at finite right-moving temperature $T_R$. As discussed at the end of that section, the  deformation would be easier to understand if the DLCQ limit commuted with the deformation. Also, we were unable to provide a proof of the fact that the identification of the $y$ coordinate in \eqref{idyT} should correspond to a right-moving temperature in the dual theory.

In this section we would like to discuss the interpretation of the identification as a right-moving temperature and of the DLCQ commuting with the deformation in the context of two-dimensional dipole theories dual to the backgrounds \eqref{nsdip} and \eqref{ssdip}. Our tool will be the fact that dipole backgrounds can be constructed via a TsT transformation from AdS, and  that we know how to relate the parameters of the TsT and the original AdS background to the deformation parameter and the temperature in the dual field theory. 

We start this section with a brief review of dipole theories. Next, we show how the dipole backgrounds \eqref{nsdip} - \eqref{ssdip} appear in the very near-horizon limit of the spacetimes dual to spacelike dipole theories, which means that they can be interpreted as DLCQs of spacelike dipole theories. Unfortunately, spacelike dipole theories at large $N$ and strong coupling do not have a nice description in terms of exact operator deformations. On the other hand, lightlike dipole theories do. We show that our dipole backgrounds  can also be thought of as DLCQs of lightlike dipole theories and we explain the relationship between the two descriptions. 

Consequently, in the particular case of dipole deformations, it is indeed true that the deformation commutes with the DLCQ. Also, one can see explicitly that the identification \eqref{idyT} does indeed correspond to the right-moving temperature in the dual theory. 

\subsection{Brief review of dipole theories}

Dipole theories are nonlocal and Lorentz non-invariant field theories. They can be obtained from the gauge theory that lives on a Dp-brane's
worldvolume via a series of transformations known as TsT (T-duality, shift, T-duality). If a Dp-brane is compactified 
on a circle of radius $R$ parametrised by a coordinate $\ty$ and $\psi$ is an angular direction in the transverse $\mathbb{R}^{9-p}$,
the TsT transformation consists of

\bi
\item a T-duality along $\ty$
\item a shift $\psi \r \psi + \l\,\ty'$, which  does not change the local geometry but introduces a twist
\item a T-duality back on the new $\ty'$ coordinate
\ei
Here $\ty'$ denotes the T-dual coordinate to $\ty$. Applying the above TsT transformation to the Yang-Mills low-energy action that lives on the Dp-brane world-volume, one  obtains a theory in which fields that were originally charged under the $R$-symmetry associated to $\psi$ rotations
are assigned a dipole vector $L^\mu_\Phi$ proportional to their R-charge 

\be
L_\Phi^\mu = 2 \pi q_\Phi\, L^\mu \;,\;\;\;\;\; L^\mu = \l\, \d^\mu_{\tilde{y}} 
\ee
Here $q_\Phi$ is the R-charge of the field $\Phi$ and $L^\mu$ is a constant vector. The Lagrangian of the dipole theory has a very simple definition in terms of a star product, which is noncommutative and nonlocal 

\be
(\Phi_1 \star \Phi_2) (x^\mu) = \Phi_1 (x^\mu- \half L_2^\mu)\, \Phi_2 (x^\mu + \half L_1^\mu)
\ee
with which one replaces all the ordinary products in the original gauge theory Lagrangian. Although the theory on the brane becomes nonlocal, it still retains several nice features, such as the fact that planar diagrams in the dipole theory are the same as in the undeformed one, up to overall phase factors which depend only on the external momenta. We believe this property should play an important role when comparing gauge theory calculations with classical gravity ones. 

The best-studied example of dipole gauge theories is the $p=3$ case, or the dipole deformations of $\N=4$ SYM. In that case, there exists an explicit Seiberg-Witten-like map \cite{berggan} which allows one to rewrite the dipole theory as  $\N=4$ SYM with an infinite number of higher-derivative terms added, all with fixed coefficients. It is quite possible that this map, which translates between ``dipole type'' of gauge invariance\footnote{The dipole field $\Phi(x)$ transforms in the $(N,\bar N)$ representation of $U(N)_{x- \half L} \times U(N)_{x+\half L}$ rather than in the adjoint of $U(N)_x$. It is this transformation rule that we refer to as ``dipole type'' of gauge invariance. } to a theory with usual gauge invariance, exists in all dimensions. It should then not be surprising that dipole backgrounds are dual to CFTs deformed by irrelevant operators which break Lorentz invariance.

The gravity backgrounds dual to dipole theories are obtained by applying the TsT transformation to the backreacted  Dp-brane geometries and then taking the usual decoupling limit $\a' \r 0 $, while keeping distances  measured in units of $\a'$  fixed. The asymptotic structure of the resulting decoupled geometries is not particularly easy to interpret from a holographic point of view.  Nevertheless, if in addition one takes a Penrose-like limit while scaling the $\l$ parameter to zero at the same time, one obtains precisely the Schr\"{o}dinger backgrounds \eqref{ddimschroed} \cite{alishganor}. Because of the infinite boost, the dipole vector $L^\mu$ becomes lightlike, and thus gravity in Schr\"{o}dinger backgrounds is dual to a lightlike dipole theory. 

The case of interest to us is the dipole deformation of the D1-D5 gauge theory, which in the IR should flow to a field theory dual to the backgrounds \eqref{nsdip} - \eqref{ssdip}. While we expect the description of this theory to be slightly more involved than the D3-brane case, we think it is reasonable to assume that the general features of the field theory - such  as the Seiberg-Witten map and the equality of planar diagrams between the deformed and the undeformed theories - will still be in place. In the following two subsections we show how the dipole backgrounds \eqref{nsdip} - \eqref{ssdip}  arise as DLCQs of spacelike or lightlike dipole theories, and outline the lessons we can learn from this construction.

\subsection{Spacelike dipole theories and DLCQ \label{spdipsec}}

To obtain the three-dimensional finite-temperature dipole backgrounds, we start from the BTZ$\times S^3$ metric, which reads

\be
ds^2 = - \frac{(\rho^2 - \rho_+^2)(\rho^2 - \rho_-^2)}{\rho^2} d\tau^2 + \frac{\ell^2\rho^2 d\rho^2}{(\rho^2 - \rho_+^2)(
\rho^2 - \rho_-^2)}
+ \rho^2 \left( R d\varphi - \frac{\rho_+ \rho_-}{\rho^2}d\tau\right)^2 + \ell^2 d \Omega_3^2
\ee
with $\varphi \sim \varphi + 2 \pi $. There is also RR three-form flux which supports this background. This black hole corresponds to a thermal ensemble in the dual $CFT_2$ characterised by temperatures $T_{L,R}$, with

\be
\rho_\pm = \pi \ell  (T_R \pm T_L)
\ee
Next, we perform a TsT transformation on this geometry, consisting of a T-duality on $\varphi$, a shift $\psi \r \psi + 2 \l \varphi'$, and then a T-duality back on $\varphi'$. The resulting metric is\footnote{The reason that there are no factors of $\a'$ in the above metric is that we are working with the decoupled geometry, in which $\a'$ has been taken to zero. A more correct way to find this metric would have been to first do the TsT and only then take the decoupling limit, but the two procedures do in fact commute.}

\bea
ds^2 &=& - \frac{(\rho^2 - \rho_+^2)(\rho^2 - \rho_-^2)}{\rho^2} d\tau^2 + \frac{\ell^2\rho^2 d\rho^2}{(\rho^2 - \rho_+^2)(\rho^2 - \rho_-^2)}
+  \frac{\rho^2}{1+ \l^2 \rho^2 \ell^2 R^2} \left(R  d\varphi - \frac{\rho_+ \rho_-}{\rho^2} d\tau\right)^2 \non \\
&&\hspace{2 cm}  + \frac{ \ell^2\s_3^2}{4(1+ \l^2 \rho^2  \ell^2 R^2)} + \frac{\ell^2}{4} d \Omega_2^2 \label{spdip}
\eea
This procedure also generates a nontrivial B-field and dilaton 

\be
B =- \frac{\l \rho^2 \ell^2 R}{2( 1 + \l^2 \rho^2 \ell^2 R^2 )} \,\left(R d\varphi- \frac{\rho_+ \rho_-}{\rho^2} d\tau 
\right)\wedge (d\psi + \cos \th d \phi) \;, \;\;\;\;\;\;
e^{- 2 \Phi} =1 + \l^2 \rho^2 \ell^2 R^2
\ee
The asymptotic structure of this spacetime is not particularly enlightening. Nevertheless, if we let $\rho_- = \rho_+= T \ell/2R$ 
and take the near-horizon limit \eqref{nearhorlim},  we obtain 

\be
 ds^2 = \frac{\ell^2}{4} \left[- \frac{r^2 d\tt^2}{T^2} + \frac{dr^2}{ r^2} + \frac{T^2}{1+ \tilde{\l}^2 T^2} \left(d\tilde{y} + \frac{rd\tt}{T^2} \right)^2 + 
\frac{\s_3^2}{1+ \tilde{\l}^2 T^2}  + d\Omega_2^2 \right]
\ee
where we have defined 

\be
\tilde{\l} = \frac{\l \ell^2}{2 }
\ee
The dilaton and the B-field become 

\be
e^{-2\Phi}= 1 + \tilde{\l}^2 T^2 \;, \;\;\;\;\;\; B = \frac{\ell^2}{4} \left[ - \frac{\tilde{\l} T^2}{1+ \tilde{\l}^2 T^2} \left(d \ty + \frac{r d\tt}{T^2} \right) \wedge \s_3\right]
\ee
The above metric is written in ten-dimensional string frame, and we have omitted the internal K3 factor. Nevertheless, in holography it is the Einstein frame metric which is 
more natural to use. The six-dimensional Einstein frame metric $g_E = e^{-\Phi} g_{S}$ is precisely the dipole background\footnote{If we interchange the roles of $\psi$ and $\phi$, we obtain the supersymmetric dipole backgrounds \eqref{ssdip} instead.} \eqref{nsdip}  with $\e_B = \tilde{\l} T$.

\be
ds^2  = \frac{\ell^2\sqrt{1+ \tilde{\l}^2 T^2}}{4} \left(  - \frac{r^2 d\tt^2}{T^2} + \frac{dr^2}{r^2} + \frac{ T^2}{1+ \tilde{\l}^2 T^2} \left(d\ty + \frac{r d\tt}{T^2} \right)^2 
+ d\Omega_{S^2}^2 + \frac{\s_3^2}{1+ \tilde{\l}^2 T^2} \right)  \label{einsdipagain}
\ee
or, in terms of our favourite coordinates $y = T \ty$, $t = T^{-1} \tt$

\be
ds^2 =  \frac{\ell^2\sqrt{1+ \tilde{\l}^2 T^2}}{4} \left(  - r^2 dt^2 + \frac{dr^2}{r^2} + \frac{ 1}{1+ \tilde{\l}^2 T^2} \left(dy + r dt \right)^2 
+ d\Omega_{S^2}^2 + \frac{\s_3^2}{1+ \tilde{\l}^2 T^2} \right) \label{einsdipyt}
\ee
Given the interpretation of the near-horizon limit as a DLCQ limit, this construction shows that the holographic dual to the dipole backgrounds \eqref{nsdip} - \eqref{ssdip} is the DLCQ of the spacelike dipole deformation of the D1-D5 gauge theory.

There are a few things to note about the metric \eqref{einsdipyt}. First, as we have anticipated, the identification of the coordinate $y$ corresponds indeed to the temperature for the right-movers in the DLCQ. Second, the interpretation of this background using holographic intuition, unlike for pure Schr\"{o}dinger backgrounds, is far from obvious. The parameter $T$ in the solution, being proportional to the right-moving temperature, characterises a state in the dual theory, whereas $\l$, being proportional to the dipole vector

\be
L^\mu = 2 \l R\, \d^\mu_{\ty}
\ee
characterises the deformation. Working in the proper holographic coordinates $\ty, \tt$, we observe that
 what we would usually identify as the source for the irrelevant operator (the coefficients of $r d\tt$ in the expression for $B$ and that of $r^2 d\tt^2$ in the metric) seems  to depend on the right-moving  temperature, and thus the state the system is in. Moreover, the radius of the geometry in Einstein frame (measured as the coefficient of the $\frac{dr^2}{r^2}$ term in the metric) also seems to depend on the temperature\footnote{Nevertheless, the central charge of the asymptotic symmetry group does not, see section \ref{asg}.}. It would be very interesting to reproduce these features using the techniques of holographic renormalization, along the lines of \cite{baltmulti}.

Note that we could have reached exactly the same background by starting from self-dual $AdS_3 \times S^3$ written in proper holographic coordinates and performing the TsT transformation on the near-horizon 
geometry directly. This shows that the dipole deformation commutes with the DLCQ or near-horizon limit. Consequently, we can think of the dipole backgrounds as either DLCQs of a spacelike dipole theory, or as dipole deformations of the DLCQ of a CFT\footnote{Provided the latter procedure can be given a precise definition.}.

While spacelike dipole theories can in principle be understood in terms of the dipole star product deformation of the D1-D5 gauge theory,  they do not have a simple interpretation as exact deformations 
of a CFT$_2$ by a $(1,2)$ operator. For this reason, we turn to studying DLCQs of lightlike dipole theories.

\subsection{Relationship to lightlike dipole theories \label{lightdip}}

Lightlike dipole theories are dipole theories in which the dipole vector $L^\mu$ is lightlike. They can be obtained via a Penrose limit of the spacelike dipole theories \cite{alishganor}. The supergravity backgrounds that they are dual to can be obtained from \eqref{spdip} by first defining lightlike ``boundary'' coordinates

\be
\hat{u} =  R\, \varphi- \tau\;, \;\;\;\;\; \hat v = R\, \varphi + \tau \;, \;\;\;\;\; \hat u \sim \hat u + 2 \pi R \;, \;\;\;\;\;
\hat v \sim \hat v + 2 \pi R
\ee
in terms of which the metric reads

\bea
ds^2 &=& \frac{d\hat u^2}{1+ \hat\l^2 \hat\rho^2} \left[ \left( \frac{\hat\rho_+ + \hat\rho_-}{2}\right)^2  - 
\frac{\hat \l^2}{4} (\hat\rho^2 - 
\hat \rho_+^2)(\hat \rho^2 - \hat\rho_-^2)\right]  + \frac{\ell^2\hat \rho^2 d\hat \rho^2}{(\hat \rho^2 - \hat\rho_+^2)
(\hat\rho^2 - \hat\rho_-^2)} + \non \\
&& + \frac{d\hat v^2}{1 + \hat \l^2 \hat \rho^2} \left[ \left(\frac{\hat \rho_+ - \hat \rho_-}{2}\right)^2 - 
\frac{\hat\l^2}{4} (\hat\rho^2 - 
\hat \rho_+^2)(\hat \rho^2 - \hat \rho_-^2)\right] + \frac{2 d\hat u d\hat v}{1+ \hat \l^2 \hat \rho^2} \left[ \frac{2\hat
\rho^2 - \hat\rho_+^2 -\hat \rho_-^2}{4} 
+\right. \non \\
&& \left. + \frac{\hat\l^2}{4} (\hat \rho^2-\hat \rho_+^2)(\hat\rho^2-\hat \rho_-^2) \right] + \frac{\ell^2 \s_3^2}{4(1+ \hat 
\l^2 \hat\rho^2)} + 
\frac{\ell^2}{4} d \Omega_2^2
\eea
Note that we have added a hat on all noncompact coordinates and horizon radii. The shift parameter is 

\be
\hat \l = \l R \ell
\ee
Next, we define the rescaled temperatures

\be
\hat T_\pm = \half (\hat\rho_+ \pm \hat\rho_-) = \pi \ell  T_{R/L}
\ee
and perform the following boost and rescalings

\be
\hat u = e^{-\g} \ty \;, \;\;\;\;\; \hat v = 2 e^\g \tt \;, \;\;\;\;\; r= \hat \rho^2 - \hat \rho_+^2 \;, \;\;\;\;\; 
\hat \l = e^{-\g } \tilde{\l} \;, \;\;\;\;\;\hat T_\pm = e^{\pm \g} T_\pm 
\ee
Taking the limit $\g \r \infty$ with $\ty,\tt,\tl,T_\pm$ fixed, we obtain 

\bea
ds^2 & = &\frac{T_+^2 d\ty^2}{1 + \tl^2 T_+^2} + \frac{2(r + 2 T_+ T_-) d\ty d\tt}{1+ \tl^2 T_+^2} + \frac{4 d\tt^2}{1+ \tl^2 T_+^2}
\left[  T_-^2 - \frac{\tl^2}{4} r (r + 4 T_+ T_-)\right] + \non \\
&& \hspace{1 cm} +\frac{\ell^2 dr^2}{4r(r+ 4 T_+ T_-)} + \frac{\ell^2 \s_3^2}{4(1+ \tl^2 T_+^2)} +\frac{\ell^2}{4} 
d \Omega_2^2
\eea
Note that we absolutely had to scale down $T_-$ in order to avoid divergences, nevertheless scaling up $T_+$
was optional (not scaling it would yield the background above with $T_+=0$, thus less interesting). The 
periodicities of the new coordinates are 

\be
\ty \sim \ty + 2 \pi  R \,e^\g \;, \;\;\;\;\; \tt \sim \tt +  \pi R  \,e^{-\g}
\ee
In the infinite boost limit the coordinate $\tt$ is no longer identified, whereas the $\ty$ circle decompactifies. Setting  $T_-=0$ and passing to Einstein frame we obtain precisely our dipole backgrounds \eqref{einsdipagain} with $\e_B = \tl T_+$, \emph{except}  that the coordinate $\ty$ is noncompact instead of compact. It is not hard to make $\ty$ compact: one simply has to take the limit

\be
R \r 0 \;,\;\;\;\;\; \g \r \infty \non
\ee
with 
\be
R_l \equiv R \,e^{\g} \label{dlcqdef}
\ee
fixed. This procedure compactifies the lightlike dipole theory onto a lightlike 
circle of radius $R_l$ and is equivalent to a DLCQ of the lightlike dipole theory. Thus, our
dipole backgrounds are also described as DLCQs of lightlike dipole theories. 

The coordinate that we use for the dipole backgrounds \eqref{nsdip} is

\be
y = T_+ \ty \;, \;\;\;\;\; y \sim y + 2 \pi R_l T_+ 
\ee   
which shows that the identification of $y$ is indeed the dimensionless temperature in the lightlike dipole theory, as 
previously advertised. 

The combination of the Penrose limit and the zero-radius limit \eqref{dlcqdef} is nothing but 
the definition of the DLCQ in the original spacelike dipole theory. The procedure presented
in the previous section is simply a different (and perhaps slightly less transparent from a field-theoretical point of view) way of implementing the DLCQ in terms of \emph{spacetime} variables; in the field theory, what we have done is the same\footnote{It is worthwhile though to show explicitly that also in this case the near-horizon limit corresponds to a DLCQ limit in the dual field theory.}. In particular, one can notice that keeping $T_{\pm}$ fixed requires
that the original left-moving temperature $\hat T_-$ be zero (and thus, we are freezing the left-movers), whereas we are concentrating on very high energy modes $\hat T_+ \r \infty$ in the right-moving side. Nevertheless, since we are simultaneously scaling the mass gap of the theory -
given by $1/R$  - to infinity, the relative energy of the high energy modes with respect to the mass gap, or $\hat T_+ R = T_+ R_l$ stays finite\footnote{This construction of lightlike dipole theories via the Penrose limit of spacelike dipole ones and the focusing of energies associated with this procedure suggests that lightlike 
dipole theories  only capture  the zero-mode dynamics in the left-moving sector of spacelike dipole theories.}. 

Note that the DLCQ limit, even for spacelike dipole theories, automatically makes the dipole vector lightlike due to the infinite boost. Consequently, there is no difference between the DLCQ of a spacelike dipole theory and that of a lightlike dipole one, except the fact that the former is easier to interpret from a brane perspective, whereas the latter is more easily dealt with from the viewpoint of exact operator  deformations. 

Note also that in the DLCQ limit $R \r 0$ the supergravity approximation breaks down. The natural description of the brane system is in the T-dual frame, in which we obtain a collection of bound D0 and D4 branes probing a twisted background. 

The lightlike background \eqref{einsdipagain} with $\ty$ not identified corresponds to a finite right-moving temperature state in the lightlike dipole theory, which can be described at large $N$ and strong coupling in terms of an exact $(1,2)$ operator deformation of the $CFT_2$ dual to $AdS_3$. We see explicitly that the DLCQ procedure (which in this case is simply a quotient along $\ty$) can be applied either before or after the deformation is performed. Moreover, the strength of the deformation is given by $\tl$ and is independent of the temperature, so in particular it equals the coefficient of the deformation that we read off in the zero-temperature limit of section \ref{3dsch}. This picture thus confirms our expectation that the computation of the correlators in the deformed theory can be performed \emph{before} taking the DLCQ limit, by using a prescription of the form \eqref{pres}.

\section{What have we learned about the Kerr black hole? \label{what}}

The motivation for the present work was the proposal of \cite{microkerr}: namely, that we can gain new insights into the mysterious microscopic side of the Kerr/CFT correspondence by embedding it into string theory. In this concluding section we would like to give an overview of the picture that appears to emerge from 
this embedding.

More concretely, the proposal of \cite{microkerr} was that one can understand the microscopic description of the five-dimensional analogue of the extreme Kerr black hole - more precisely, of the extreme Myers-Perry black hole with equal angular momenta - by viewing it as a particular representative in a one-parameter family of extremal, non-supersymmetric, charged and rotating five-dimensional black holes. 
The black holes are parametrised by their charge $Q$ and left-moving angular momentum $J_L$ with $Q^3 \leq J_L^2$. If we keep $J_L$ fixed as we vary $Q$ and concentrate on the near-horizon limit of this family of black holes, the near horizon geometries continuously interpolate between that of  the extremal Myers-Perry black hole at $Q =0$ and $AdS_3 \times S^2$ when $Q^3 = J_L^2$. This can be summarized by the following picture\footnote{Note that, contrary to any common-sense conventions, $Q$ decreases along the direction of the arrow.}

\medskip

\begin{figure}[h]
\centering
\includegraphics[scale=0.50]{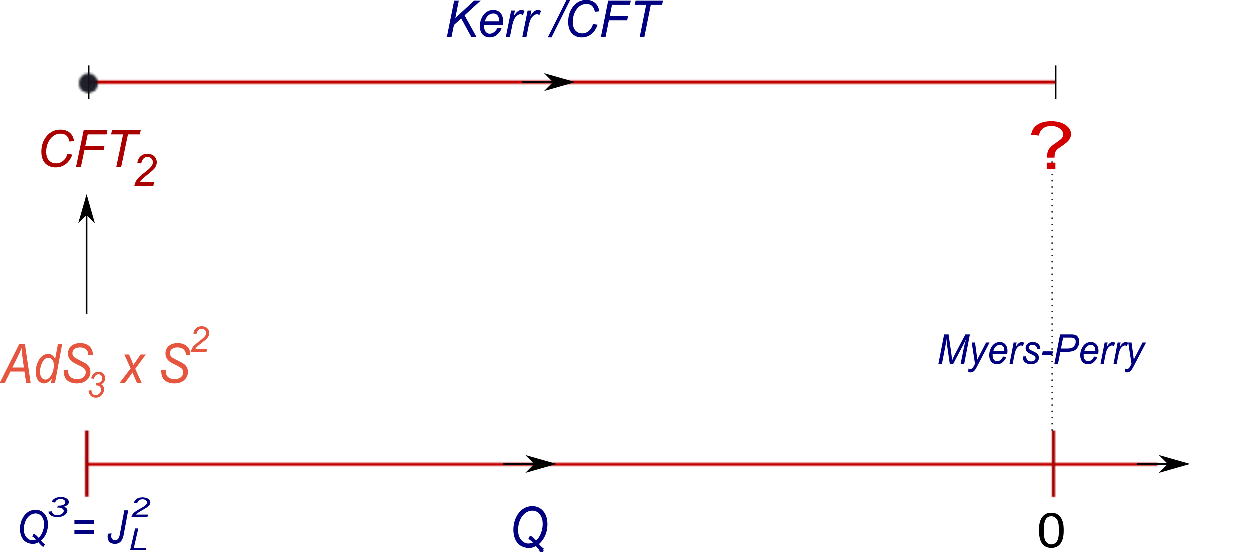}
\caption{A family of $5d$ extremal charged, rotating black holes with the same value of $J_L$ and varying $Q$, which interpolates between the Myers-Perry black hole at $Q=0$ and the maximal value $Q^3 = J_L^{2}$. At precisely this point, the near horizon geometry becomes $AdS_3 \times S^2$ and the dual CFT description is well understood.}
\end{figure}

\medskip

\noindent All the black holes in the family allow for an asymptotic symmetry group analysis analogous to that in \cite{kerrcft}, which leads to the conjecture that they are all states in a ``CFT\footnote{Throughout this section we will use quotes whenever we refer to CFTs whose presence is signaled by an asymptotic symmetry group analysis but are not yet proven to be standard CFTs. }'' with central charge 

\be
c_L = 6 J_L \;, \;\;\;\;\; T_R = \frac{1}{2\pi} \sqrt{1 - \frac{Q^3}{J_L^2}} \label{parambh}
\ee 
and are characterised by the right-moving temperature $T_R$ given above. The asymptotic symmetry group analysis does not make it clear whether they correspond to different thermal states in the \emph{same} ``CFT'' of central charge $6 J_L$, or whether one has a one-parameter family  of ``CFT''s with equal central charges, parametrized by $Q$. 

The point of view taken in \cite{microkerr} was that we are dealing with a one-parameter family of ``CFT''s. The central observation of that paper is that precisely at the maximal point $Q^3 = J_L^2$ the ``CFT'' is a CFT$_2$ in the usual sense of the word, with central charge $6J_L$ and at \emph{zero  temperature}. Nevertheless, precisely at this point the near-horizon geometry becomes $AdS_3$, so the example \cite{microkerr} found is a slightly trivial representative of a Kerr CFT. 

Given that we have a continuous family of near horizon geometries, one may nevertheless ask whether one can understand the ``CFT'' representatives which are $\O(\e)$ away from the maximal point,  where $\e \propto T_R$, by using the holographic dictionary in $AdS_3$. Such a task is difficult to achieve in five dimensions, because the passage from the exactly maximal to the near-maximal near-horizon geometry involves a change in topology similar to the one in \cite{m2m5}. A simple solution \cite{microkerr} is to require that the black hole in question admit a six-dimensional embedding, in which case the near-maximal geometry \emph{can} be studied using the perturbative AdS/CFT dictionary. Thus, we obtain the  picture presented in Figure 2 below.

\bigskip

\begin{figure}[h]
\centering
\includegraphics[scale=0.60]{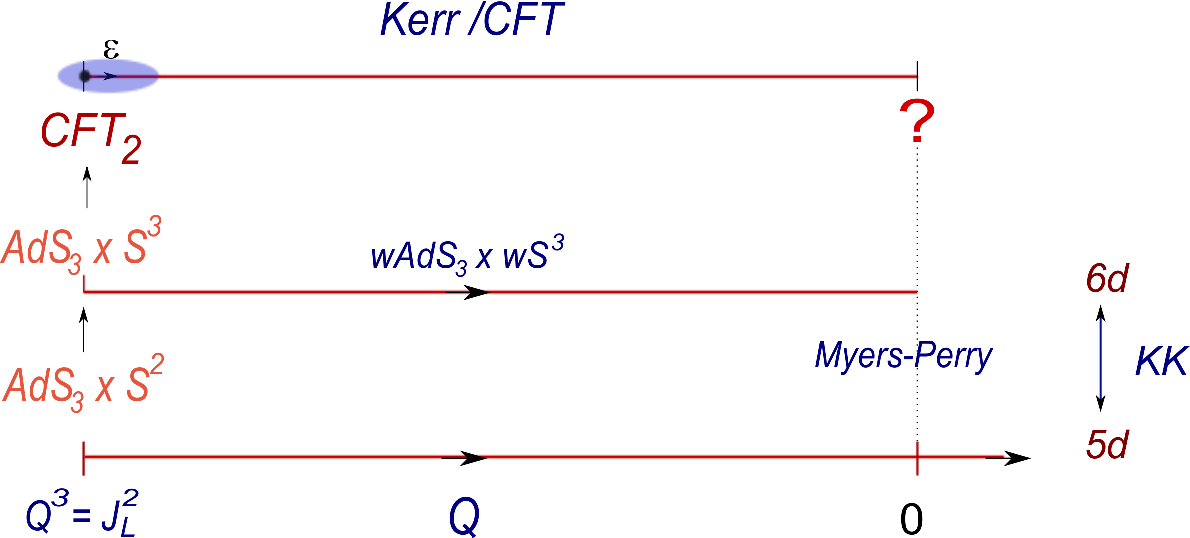}	\label{fig:6dkerr}
	\caption{The uplift of the one parameter family of black holes to six dimensions. Now the dual field theory can be understood at $\O(\e)$ away from maximality by using AdS/CFT. Here $\e \propto T_R$ labels both the temperature and the coefficient of the operator deformation.  }
\end{figure}

\medskip

\noindent The result of the holographic analysis is that $\O(\e)$ away from maximality the system acquires a right-moving temperature ($T_R$), but also an \emph{irrelevant}, Lorentz non-invariant, $(1,2)$ operator is turned on. Deforming a CFT by an irrelevant operator not only breaks conformal invariance, but even worse, in general it does not lead us to any sort of  well-defined theory, because the perturbation is non-renormalizable. At higher order in $\e$ we would expect even more irrelevant operators to be turned on, so we are faced with a big puzzle - which is  how to make sense of the ``Kerr CFTs'' away from the strictly maximal limit.

The resolution  proposed in the present article was inspired by \cite{nr} and is that the $(1,2)$ deforming operator should not be thought of as an \emph{irrelevant} operator with respect to the $2d$ conformal group, but rather as a \emph{marginal} operator with respect to the Schr\"{o}dinger group\footnote{Note that the symmetry preserved by the deformation strongly
constrains the form of the operators generated by the deformation at higher order. In turn, this has the 
potential to render conformal perturbation theory  well defined - at least at large N and strong coupling.} of nonrelativistic conformal transformations in zero spatial dimensions, which is  $SL(2,\mathbb{R}) \times U(1)_{null}$. With this interpretation, it does make sense to add the perturbing operator to the CFT action, and we find ourselves at a nonrelativistic conformal fixed point. 
This picture is corroborated by the fact that there exists a \emph{different maximal limit} of the six-dimensional family of near-horizon geometries, which does not yield $AdS_3 \times S^3$, but rather $Schr_3 \times S^3$. Three-dimensional Schr\"{o}dinger spacetimes are the paradigmatic spacetimes dual to non-relativistic CFTs. Note that in this picture the deformation does not vanish even at strict maximality, and the reason that we were previously obtaining a CFT$_2$ was that we were taking an infrared limit in addition to the zero-temperature limit. Going away from the maximality does not require  any additional operators to be turned on but, as we argued in the present paper, the only parameter that changes from one black hole to the next is the \emph{temperature} in the dual nonrelativistic CFT. Consequently, the extremal Myers-Perry black hole with equal angular momenta corresponds to a state of right-moving temperature $T_R = \frac{1}{2\pi}$ in a non-relativistic CFT.

\bigskip

\begin{figure}[h]
\centering
	\includegraphics[scale=0.60]{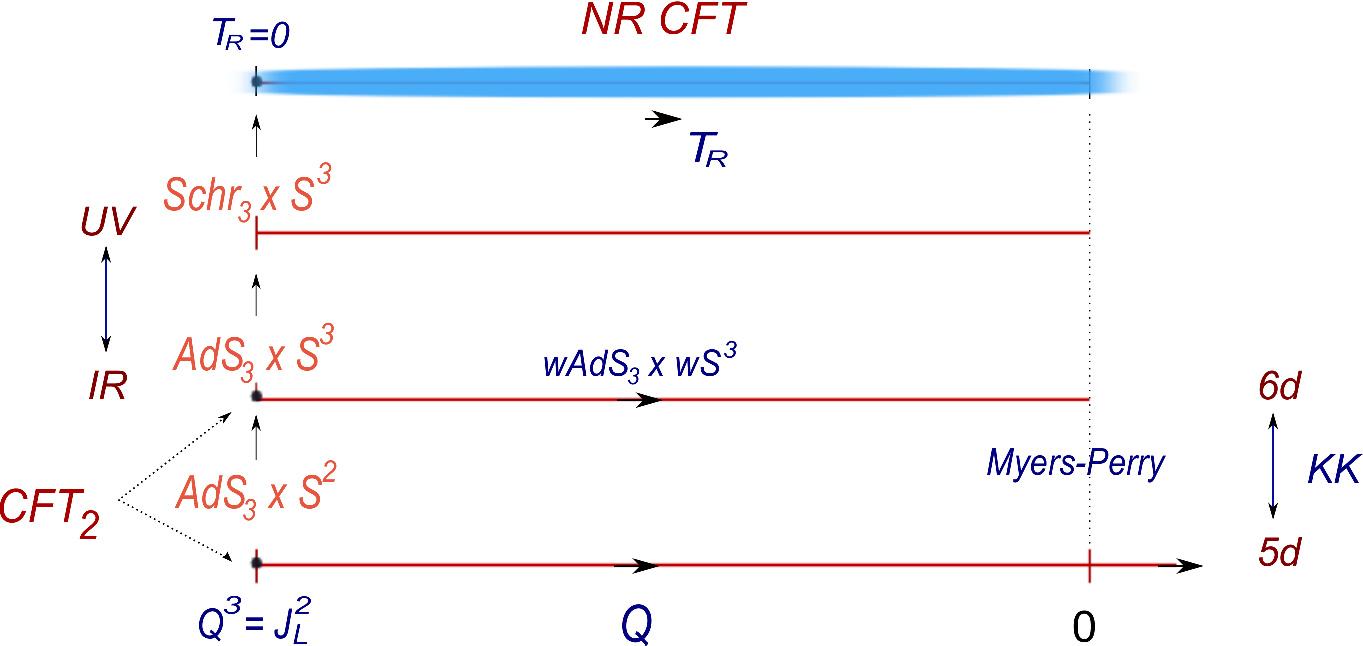}	\label{schrkerr}
	\caption{A different maximal limit of the six-dimensional geometries shows that all the black holes should be thought of as states of different temperatures inside the \emph{same} field theory, which is a nonrelativistic CFT.}
\end{figure}

\medskip

\noindent This picture is of course not complete without a definition of the nonrelativistic CFT in question. In this article we have proposed an effective definition of this CFT, which should hold at large $N$ and strong coupling, namely in the regime where the dual classical supergravity description is valid. The prescription is to simply define the nonrelativistic CFT as the theory obtained by adding a \emph{single} exactly marginal operator with respect to nonrelativistic scaling to the original CFT$_2$ action. Correlation functions in the deformed theory should then be computable in terms of the ones in the original CFT by using conformal perturbation theory. 

If the above prescription is correct, then all features of spacetime physics should be reproducible from the field theory perspective, including the Virasoro asymptotic symmetry group and its central extension. Note that the global symmetries of the non-relativistic CFTs are precisely $SL(2,\mathbb{R}) \times U(1)$, which means that they may constitute an example of a class of theories to which the results of \cite{stromhof} apply.   

This article presents a number of loose ends. The most important is probably our inability to provide a proof that the intermediate near-horizon geometries are finite right-moving temperature versions of the maximal one, although we present plenty of circumstantial evidence. Another claim whose proof needs to be completed is that of the absence of multitrace operators from the deforming action, i.e. that the non-relativistic CFT is defined by deforming the original CFT by a \emph{single} $(1,2)$ operator. Yet another fact to understand is whether the coefficient of the deforming operator could depend on the temperature. Our parallel analysis of dipole backgrounds has indicated that this is not the case in that specific example, but we do need a much better understanding of the holographic dictionary in asymptotically Schr\"{o}dinger spacetimes in order to answer these questions. An important formal issue to prove is that conformal perturbation theory in the deforming Lorentz-breaking operator is well-defined. Finally, all the geometries that enter the discussion correspond to DLCQs of the dual field theories, which need to be much better understood.  

We should also mention that the picture presented in this section is not an entirely accurate account of the results we have proven in the paper, due to the fact that we were using a slightly different - but equivalent - description of the black holes\footnote{The difference is explained in section \ref{holoint}.}. More concretely, in the description we actually used, the family of black holes is specified by fixing the charge $Q$ and letting the angular momentum $J_L$ be the interpolating parameter, $J_L^2 \geq Q^3$. The ``CFT'' descriptions now all have fixed central charge $c = 6 Q^2$. All results about the interpretation of the maximal limit and the intermediate geometries still hold (and even more so), and the resulting picture is

\bigskip

\begin{figure}[h]
\centering
\includegraphics[scale=0.60]{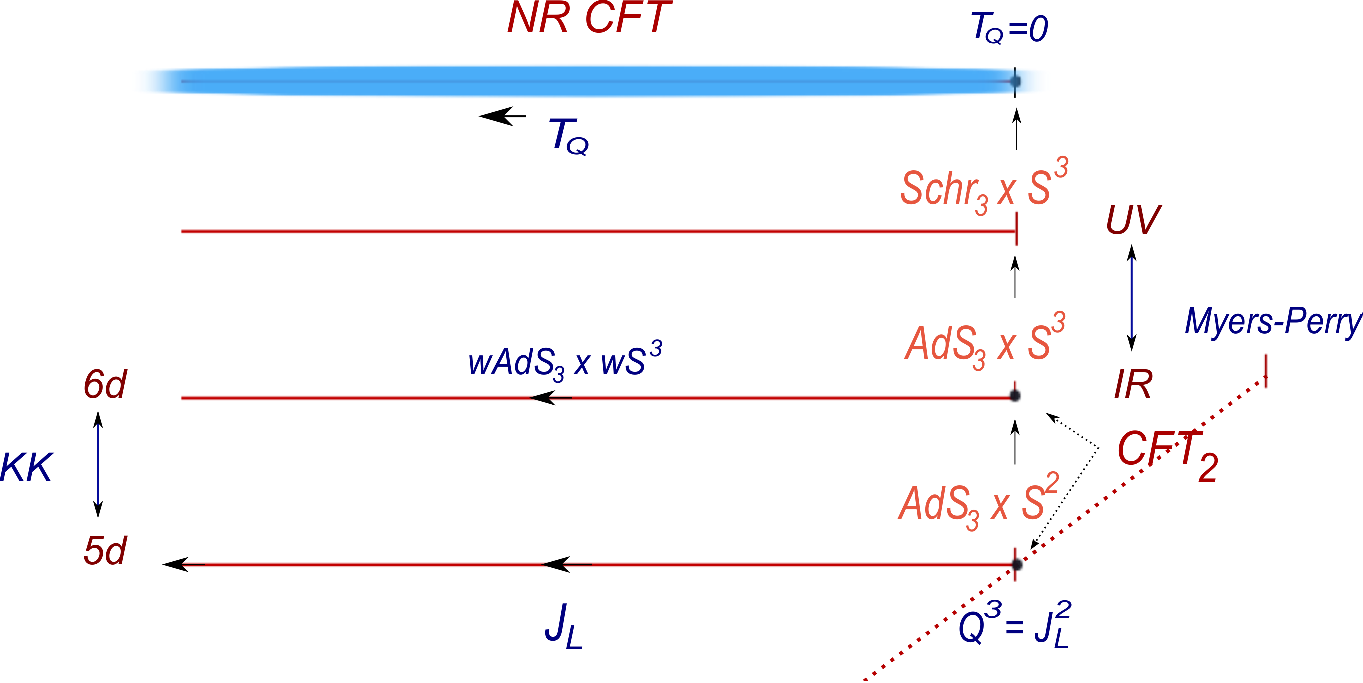}	
	\label{qsqkerr}
\caption{Another descr‬iption of the family of black holes, now seen as having fixed $Q$ and varying $J_L$, as different thermal states with temperatures $T_Q$ in a nonrelativistic CFT. The lower bound on the angular momentum is $J_L^2 \geq Q^3$, so the Myers-Perry black hole is never a part of this family.}	
\end{figure}

\medskip

\noindent
As explained in section \ref{holoint}, the advantages of this description lie in the fact that the non-relativistic CFT is now a deformation of the $(4,4)$- supersymmetric  D1-D5 CFT, which is very well understood, and that the quotients that act on the geometry have a clear interpretation even away from maximality. While one can consider black holes which have $Q$ small compared to $J_L$, this description breaks down in the exact Myers-Perry limit, and we have to understand the ``Kerr/CFT -like'' description which, as explained in section \ref{d1d5rev}, is slightly more involved. We do expect though that the main view of this paper will carry over to this case. 

A very interesting question is whether the analysis of the present article also applies to the construction in \cite{compwei}, which presents a set of geometries that interpolate between $AdS_3 \times S^2$ and the extremal $4d$ Kerr black hole ($\times S^1$). It would also be very interesting to understand whether, more generally, families of black holes which contain a representative whose near horizon region is of the pinching orbifold type admit a description as thermal states in a non-relativistic CFT.  It would also be interesting to explore the relationship with the EVH/CFT correspondence of \cite{sheikh}.

\bigskip

\noindent\textbf{Acknowledgements}

\medskip

\noindent We would like to thank I.~Bena, N.~Bobev, J.~de Boer, J.~Gauntlett, S.~Ross, J.~Simon and K.~Skenderis for interesting discussions.  We are especially grateful to
 G.~Comp{\`e}re, K.~Papadodimas, B.~van Rees and A.~Strominger for many insightful discussions
and useful comments on the manuscript. We would also like to thank the ``Centro de Ciencias de Benasque Pedro Pascual'', where part of this work was completed, for hospitality and a wonderful work environment. M.G. would also like to thank the  Aspen Center for Physics for hospitality and the 
National Science Foundation for partial support, under Grant No. 1066293, during her stay in Aspen. The research of S.E. is supported by the Netherlands Organization for Scientific Research (NWO) under a Rubicon grant.  The work of S.E. and M.G. was also supported by the  ANR grant 08-JCJC-0001-0, and by the ERC Starting Independent Researcher Grant 240210 -String-QCD-BH.

\appendix

\section{Hodge duals}

For future reference we list herein the Hodge duals of various three-forms of interest with respect to the metric \eqref{bck}

\bea
\star(\s_1 \wedge \s_2 \wedge \s_3) & = & \frac{1}{2 \sqrt{1-\e_g^2}}  w_+ \wedge w_- \wedge (w_3 + \e_g \s_3) \non \\
\star(w_+ \wedge w_- \wedge w_3) & = & \frac{2}{ \sqrt{1-\e_g^2}}  \s_1 \wedge \s_2 \wedge (\s_3 + \e_g w_3) \non \\
\star(\s_1 \wedge \s_2 \wedge w_3) & = & -\frac{1}{2 \sqrt{1-\e_g^2}}  w_+ \wedge w_- \wedge (\s_3 + \e_g w_3) \non \\
\star(w_+ \wedge w_- \wedge \s_3) & = & -\frac{2}{ \sqrt{1-\e_g^2}}  \s_1 \wedge\s_2 \wedge (w_3 + \e_g \s_3) 
\eea
Therefore

\bea
\star H & = & \frac{\ell^2}{\sqrt{1-\e_g^2}} \left[ (h - \g \e_B \e_g) \left( \s_1  \wedge \s_2 \wedge \s_3 + \half w_+ \wedge
w_- \wedge w_3\right) + \right. \non \\
&& \left. \hspace{3 cm} + (h \e_g - \g \e_B) \left( \s_1  \wedge \s_2 \wedge w_3 + \half w_+ \wedge
w_- \wedge \s_3\right)  \right]
\eea

\section{Killing spinors of $3d$ Schr\"{o}dinger spacetimes \label{killsp}}

In this appendix we show, using a Killing spinor analysis, that the $SL(2,\mathbb{R})_L \times SU(2)_R$ invariant Schr\"{o}dinger backgrounds preserve $(0,4)$ Poincar\'e supersymmetry, whereas the $SL(2,\mathbb{R})_L \times SU(2)_L$ invariant ones preserve $(4,0)$ superconformal symmetry. Our results perfectly parallel those obtained in \cite{gaunt} for the case of five-dimensional Schr\"{o}dinger backgrounds. We first perform the analysis for the dipole backgrounds $(\l_g =0)$, and then for the self-dual Schr\"{o}dinger backgrounds $(\l_g = 2 \l_B)$. We set the AdS length $\ell=2$ throughout this section.

\subsection{Dipole backgrounds}

The dipole type Schr\"{o}dinger backgrounds have $\l_g =0$. Their metric and three-form fields read

\be
ds^2 = - \l_B^2 r^2 d\tt^2 + \frac{dr^2}{r^2} + 2 r d\tt d\ty + d\th^2 + \sin^2 \th d\phi^2 + (d\psi + \cos\th d\phi)^2  
\ee

\be
F_3 = F_3^{(0)} = \s_1 \wedge \s_2 \wedge \s_3 + \a\, dr \wedge d\tt \wedge d\ty
\ee

\be
 H_3 =\l_B \,d (r d\tt \wedge \s_3) = \l_B (dr \wedge d\tt \wedge \s_3 + \s_1 \wedge \s_2 \wedge r d\tt)
\ee
Changing the sign of the parameter $\a = \pm 1$ allows us to switch between the  $SL(2,\mathbb{R})_L \times SU(2)_R$ invariant backgrounds for $\a=1$ and the  $SL(2,\mathbb{R})_L \times SU(2)_L$ invariant ones for $\a=-1$, provided we simultaneously interchange the angles $\phi$ and $\psi$. Note that $F_3$ remains self-dual if we perform the two operations simultaneously.

We choose the vielbein to be

\be
e^+ = r d\tt \;,\;\;\;\; e^- = d\ty - \frac{ \l_B^2}{2} \, r d\tt \;, \;\;\;\;\; e^2 = \frac{dr}{r} \non
\ee
\be  e^3 = d \th \;, \;\;\;\;\; e^4 = \sin \th d\phi \;, \;\;\;\;\; e^5= d \psi + \cos\th d\phi
\ee
in terms of which the field strengths are

\be
F_3 = \a \, e^+ \wedge e^- \wedge e^2 + e^3 \wedge e^4 \wedge e^5
\ee
and

\be
H_3 = \l_B \, (e^+ \wedge e^3 \wedge e^4 - e^+ \wedge e^2 \wedge e^5)
\ee
The spin connection $ \om_{\mu a b} = e_{\nu a} \nabla_\mu e^{\nu}{}_b $ reads 

\be
\om_{+-} = \frac{dr}{2 r} = \frac{e^2}{2} \;, \;\;\;\;\; \om_{+2} = \frac{2 d\ty- 3\l_B^2 r d\tt}{4} = \frac{e^- - \l_B^2 e^+}{2} \;, \;\;\;\;\; \om_{-2} = \frac{r d\tt}{2} = \frac{e^+}{2}  \non
\ee
\be
\om_{3 4} = \frac{d\psi - \cos \th d\phi}{2} = \frac{e^5 - \cot \th \, e^4}{2} \;, \;\;\;\;\; \om_{35} = \frac{\sin \th d \phi}{2} = \frac{e^4}{2} \;, \;\;\;\;\; \om_{45}= - \frac{d\th}{2} = - \frac{e^3}{2}
\ee
Both the dilaton and axion are zero. The type IIB supersymmetry variations read \cite{benaroib} 

\be
\d \l = -\frac{1}{4} \, \slashed{H}_3 \, \s^3 \ve - \frac{1}{4} \slashed{F}_3 \, \s_1 \ve 
\ee

\be
 \d \psi_M = \nabla_M \varepsilon  - \frac{1}{8} \G^{PQ} H_{MPQ}\, \s^3\ve + \frac{1}{8} \slashed{F}_3 \G_M \, \s^1 \ve
\ee
Here $\ve$ is a doublet of $10 d$ Majorana-Weyl spinors of negative chirality 

\be
\G^{11} \ve_{1,2} = - \ve_{1,2}
\ee
and $\slashed{F}_3$, $\slashed{H}_3$ are given by

\be
\slashed{F}_3 = \G^{MNP} F_{MNP} = \a \G^{[+-]2} + \G^{345}\, \;\;\;\;\; \slashed{H}_3=  \l_B \G^{+34} (1 + \G^{2345})
\ee
where square brackets denote anti-symmetrization. The spinor covariant derivative is

\be
\nabla_\mu \e = (\p_\mu + \Omega_\mu) \e \;, \;\;\;\;\; \Omega_\mu = \half\, \om_{\mu ab} \S^{ab} \;, \;\;\;\;\; \S^{ab} = \frac{1}{4} [\G^a,\G^b] 
\ee
The dilatino variation reads

\be
\a \,\G^{[+-]2} (1+ \a \,\G^{(6)}) \, \s_1 \ve + \l_B \G^{+34} (1+ \G^{2345})\, \s_3 \ve =0 \label{dilv}
\ee
Multiplying the above equation by $\G^+\s_1 $ we obtain that

\be
\G^{+} (1+ \a \, \G^{(6)}) \, \ve =0
\ee
Next, we multiply \eqref{dilv} by $(1-\a \G^{(6)})$ and use the equation above to find the following projection constraints

\be
(1+ \a\, \G^{(6)}) \, \ve =  \l_B \G^+ (1+ \G^{2345}) \,  \ve = 0 \label{prjctr}
\ee
Next, we turn to the gravitino variation. The $M= y$ component is the same equation as in AdS

\be
\p_y \ve + \frac{1}{4} \G^{+2} (1- \a \,\s_1) \ve =0
\ee
This automatically implies that $\p_y^2 \ve =0$. We will find it convenient to split the doublet $\ve$ according to its eigenvalue with respect to $\s_1$. Defining the projectors

\be
P_{\pm} \equiv \half \, (1 \pm \s_1)
\ee
the above equation splits into

\be
\p_y \ve_\a =0 \;, \;\;\;\;\; \p_y \ve_{-\a} + \half \, \G^{+2} \ve_{-\a} =0
\ee
with solution 

\be
\ve_{-\a} = \hat \ve_{-\a} - \frac{y}{2} \, \G^{+2} \hat \ve_{-\a}\;, \;\;\;\;\; \p_y \hat \ve_{-\a} =0
\ee
The $r$ component of the gravitino variation yields 

\be
r \p_r \ve + \frac{1}{4} \, \G^{[+-]} (1+ \a \,\s^1) \ve - \frac{\l_B}{4} \G^{+5} \,\s^3 \ve =0
\ee
This decomposes into the following equations

\be
r \, \p_r \ve_\a + \half \G^{[+-]} \ve_\a - \frac{\l_B}{4} \G^{+5} \s^3 \hat \ve_{-\a}=0 \label{epreq}
\ee

\be
r \p_r \hat \ve_{-\a} - \frac{\l_B}{4} \G^{+5} \s^3 \ve_\a =0 \;\;\;\; \Rightarrow \;\;\;\; \G^+ \p_r \hat \ve_{-\a}=0
\ee
The last equation ensures that the first one can be integrated. We obtain

\be
\hat \ve_{-\a} =  \ve_{-\a}^0 + \frac{\l_B}{2} \G^{+5} \s_3 \,\ve_\a^0  \;, \;\;\;\;\; \ve_\a = \ve_\a^0 + \frac{\l_B}{2} \G^{+5} \s^3  \ve_{-\a}^0
\ee
where $ \ve_{-\a}^0$ is an $r$-independent constant of integration, while $\ve_\a^0$ solves the homogenous version of equation \eqref{epreq}, which is the same equation that we would have had in $AdS_3$. 

Next, we solve for the angular components of the gravitino equation. We obtain

\be
\p_\th \ve - \frac{1}{4} \G^{45} (1-\s^1) \ve +\frac{\l_B}{4} \G^{+4} \s^3 \ve =0 
\ee

\be
\p_\phi \ve + \frac{1}{4} (\sin \th \G^{35} - \cos \th \G^{34}) (1-\s^1) \ve + \frac{\l_B}{4} (\cos \th \G^{+2} - \sin \th \G^{+3}) \s^3 \ve =0
\ee

\be
\p_\psi \ve + \frac{1}{4} \G^{34} (1+ \s^1) \ve + \frac{\l_B}{4} \G^{+2} \s^3 \ve =0
\ee
These equations are solved as follows: let us take the $\th$ equation for example. We project it onto left and right $\s^1$ chirality and write everything in terms of $ \ve^0_{\pm \a}$. The vanishing of the $y$-dependent part of the equation yields

\be
\G^+ \left(\p_\th  \ve^0_{-\a} - \frac{1}{4} (1+\a) \G^{45}  \ve^0_{-\a} \right) =0
\ee
which in turn implies that

\be
\p_\th \ve_\a^0- \frac{1}{4} (1-\a)\, \G^{45} \ve_\a^0 =0
\ee
and

\be
\p_\th \ve^0_{-\a} - \frac{1}{4} (1+\a) \G^{45}  \ve^0_{-\a}=0
\ee
which is clearly consistent with the previous equation. Solving also for the $\phi, \psi$ component, we obtain

\be
\p_\phi \ve_{\pm\a}^0 + \frac{1}{4} (1\mp\a) (\sin \th \G^{35} - \cos \th \G^{34}) \ve_{\pm\a}^0 + \frac{\l_B}{4} (1\mp \a) \cos \th\, \G^{+2}  \s^3  \ve_{\mp\a}^0 =0 
\ee

\be
\p_\psi \ve_{\pm\a}^0 + \frac{1}{4} (1 \pm \a) \G^{34} \ve_{\pm\a}^0  + \frac{\l_B}{4} (1\mp\a) \G^{+2} \s^3  \ve_{\mp\a}^0 = 0 
\ee
Given that $\ve^0_\a$ is $r$-dependent whereas $\ve_{-\a}^0$ is not by assumption, the above equations imply an additional projection 

\be
 (1+\a) \G^+ \ve_\a^0 = (1-\a) \G^+ \ve_{-\a}^0 =0 \;\;\; \Rightarrow \;\;\; \G^+ \ve_+^0 = \G^+ \ve_+ =0 \label{addprojc}
\ee
irrespective of the value of $\a$. The only remaining equation to check is the $t$ component of the gravitino variation. We find

\be
r^{-1} \p_t \ve + \frac{1}{2} \G^{-2} P_\a \ve  - \frac{\l_B^2}{4} \G^{+2} (1+ P_{-\a}) \ve  - \frac{\l_B}{4} \G^{34} (1+ \G^{2345}) \s^3 \ve =0 \label{delpsit}
\ee
It will now be useful to distinguish between the two cases $\a = \pm 1$, as the projection conditions \eqref{addprojc} lead to very different conclusions in each case. 

\subsubsection*{The $\a=1$ case}

Multiplying \eqref{delpsit} by $\G^+$ and using the fact that $\G^+ \ve_+ =0$, we find that

\be
\ve_+ =0 \;\;\; \Rightarrow \;\;\;\; \ve_+^0 = \G^+  \ve_-^0 =0 
\ee
All these projection conditions imply that

\be
\ve_- = \hat \ve_-= \ve_-^0 
\ee
From \eqref{delpsit} there also follows that $\p_t \ve_-=0$.

To recapitulate, for the $SL(2,\mathbb{R})_L \times SU(2)_R$ invariant background corresponding to $\a =1$ we have found that there is one Killing spinor, $\ve_-$, which satisfies

\be
\p_t \ve_- = \p_y \ve_- = \p_r \ve_- = \p_\psi \ve_- = 0 
\ee 

\be
\p_\th  \ve_- - \half \G^{45}  \ve_-=0 \;, \;\;\;\;\;\p_\phi  \ve_- + \frac{1}{2} (\sin \th \G^{35} - \cos \th \G^{34})  \ve_- =0 
\ee
and the projection conditions

\be
\G^+ \ve_- = (1 + \G^{(6)}) \ve_- =0 \label{projcond}
\ee
In addition, the $K3$ equations imply that $\ve$ is a covariantly constant spinor on the internal manifold, satisfying $\nabla_i \ve =0 $. Therefore we can decompose $\ve$ into a six-dimensional and a four-dimensional internal part

\be
\ve_- = \xi_-^I \otimes \eta_I 
\ee  
where the index $I = \{1,2 \}$ labels the two chiral, covariant constant spinors on K3. The projection conditions \eqref{projcond} act only on the six-dimensional spinor $\xi$. A generic $6d$ spinor has eight complex components. Imposing the three projections $P_-$ and \eqref{projcond} on it we reduce the number of components to two real ones. Given that $K3$ has two chiral spinors, this yields a total of four real supersymmetries, as expected. Note that the Killing spinors do not depend on the noncompact coordinates, as expected, given that they generate Poincar\'{e} rather than superconformal symmetries.
The results of our Killing spinor analysis agree with those found in \cite{baltbob}.

\subsubsection*{The $\a=-1$ case}

In this case, the projection condition \eqref{addprojc} implies that $\ve_+$ does not depend on $y$. All the equations previously written still hold, with the simplification that

\be
\ve_- = \ve_-^0 \;, \;\;\;\;\; \ve_+  = \hat \ve_+ = \ve_+^0 + \frac{\l_B}{2} \G^{+5} \s^3 \ve_-  
\ee
The differential equations obeyed by $\ve^0_+$ simplify to

\be
\p_y \ve_+^0 = \p_r \ve_+^0 = \p_\th \ve_+^0 = \p_\phi \ve_+^0 =0\;, \;\;\;\;\; \p_{\psi} \ve_+^0 + \half \G^{34} \ve_+^0 =0 
\ee
and of $\ve_-$ 

\be
\p_y \ve_- = \p_\psi \ve_-=0 \;, \;\;\;\;\; r \p_r \ve_- + \half \G^{[+-]} \ve_- =0
 \ee

\be
\p_\th \ve_- - \half \G^{45} \ve_- =0 \; \;\;\;\;\; \p_\phi \ve_- + \half (\sin \th \G^{35} - \cos \th \G^{34}) \ve_- =0
\ee
Thus, $\ve_+^0$ depends non-trivially on $\psi$, whereas $\ve_-$ depends non-trivially on $(r,\th, \phi)$. Using this fact and plugging into \eqref{delpsit}, we obtain

\be
\p_t \ve_+^0 =0 \;, \;\;\;\;\; (1+ \G^{2345}) \ve_+^0 =0
\ee

\be
r^{-1} \p_t \ve_- + \half \G^{-2} \ve_- =0 \;, \;\;\;\;\; \G^{+-} \ve_- = (1+ \G^{2345})\ve_-
\ee
The Killing spinor $\ve_- (r,t,\th,\phi)$ is precisely the same as the generator of left-moving superconformal symmetries in $AdS_3$. The last condition implies \eqref{prjctr} for $\ve_-$ when $\a = -1$. Consequently, we have preserved four superconformal symmetries on the left. As far as the right-moving side is concerned, we have one potential Killing spinor candidate, $\ve_+^0$, which satisfies

\be
\G^+ \ve_+^0 = (1+ \G^{2345}) \ve_+^0 =(1- \G^{(6)}) \ve_+^0 =0 
\ee
These conditions are in fact incompatible, so there is no supersymmetry on the right.

\subsection{Self-dual backgrounds}

In the self-dual case with $\ell=2$, the metric reads
\be
ds^2 = - 3 \l^2 r^2 d\tt^2 + \frac{dr^2}{r^2} + 2 r d\tt d\ty + d\th^2 + \sin^2 \th d\phi^2 + (d\psi + \cos\th d \phi + 2 \l r d\tt)^2
\ee
where we have set $\l_g = 2 \l_B = 2 \l$ and only $F_3^+ $ flux is turned on

\be
F_3 = \s_1 \wedge \s_2 \wedge \s_3 + dr \wedge d\tt \wedge d\ty + \l (dr\wedge d\tt \wedge \s_3 + \s_1 \wedge \s_2 \wedge r d\tt)
\ee
The vielbein is chosen to be

\be
e^+ = r d\tt \;, \;\;\;\;\; e^- = d\ty - \frac{3 \l^2}{2} r d\tt \;, \;\;\;\;\; e^2 = \frac{dr}{r}\;, \;\;\;\;\; e^3 = d \th \non
\ee

\be
e^4 = \sin \th d \phi \;, \;\;\;\;\; e^5 = d\psi + \cos \th d\phi + 2 \l r d\tt
\ee
and yields the spin connection

\be
\om_{+-} = \frac{dr}{2r} \;, \;\;\;\;\; \om_{+2} = \frac{d\ty}{2} - \frac{\l^2}{4} r d \tt + \l (d\psi + \cos\th d\phi)\;, \;\;\;\;\; \om_{+5} = \frac{\l dr}{r}\;, \;\;\;\;\;\om_{-2} = \frac{r d\tt}{2}  \non
\ee

\be
\om_{25} = - \l r d\tt\;, \;\;\;\;\;\om_{34} = \frac{d\psi-\cos\th d\phi}{2} + \l r d\tt\;, \;\;\;\;\;\om_{35} = \frac{ \sin \th d \phi}{2} \;, \;\;\;\;\; \om_{45} = - \frac{ d\th}{2}
\ee
The spinor equations are now  simply

\be
\slashed{F}_3 \s^1 \ve =0 \;, \;\;\;\;\; \nabla_M \ve + \frac{1}{8} \slashed{F}_3 \G_M \s^1 \ve =0
\ee
with 

\be
\slashed{F}_3 = \G^{[+-]2} (1+ \G^{(6)}) - \l \G^{+34} (1 - \G^{2345})
\ee
Using the same procedure as before, we find the projections

\be
(1+ \G^{(6)}) \ve = \G^+ (1- \G^{2345}) \ve  =0
\ee
The $\psi$ and $\th$ components of the gravitino equation imply that

\be
\p_\th \ve_+ - \frac{\l}{4} \G^{+4} \ve_+ =0 \;, \;\;\;\;\;\; \p_\psi \ve_- + \frac{3 \l}{4} \G^{+2} \ve_- =0
\ee
Requiring that the Killing spinor solution be periodic with respect to the above angular coordinates, we find the constraints

\be
\G^+ \ve_+ = \G^+ \ve_- =0 
\ee
The $r$ and $y$ components of the gravitino equations imply that

\be
\p_r \ve_- = \p_y \ve_\pm =0 \;, \;\;\;\;\; r \p_r \ve_+ + \half \G^{[+-]} \ve_+ =0
\ee
The remaining angular component equations are the same as the ones in $AdS_3$ and read

\be
\p_\th \ve_+ = \p_\phi \ve_+ =0 \;, \;\;\;\;\; \p_\psi \ve_+ + \half \G^{34} \ve_+ =0
\ee

\be
\p_\psi \ve_- =0 \;, \;\;\;\;\; \p_\th \ve_- - \half \G^{45} \ve_- =0 \;, \;\;\;\;\; \p_\phi \ve_- + \half (\sin \th \G^{35} - \cos \th \G^{34}) \ve_- =0 
\ee
Finally, the $t$ component of the gravitino equation reads 

\be
\p_t \ve + \frac{r}{4} \G^{-2} (1+ \s^1)\ve  + \frac{\l r}{4} \G^{34} (2 + \s^1) \ve - \frac{\l r}{4} \G^{25} (2+ \s^1) \ve =0 \label{teqn}
\ee
Projecting, we find that

\be
\p_t \ve_- + \frac{\l r}{4} (\G^{34} - \G^{25}) \ve_- =0
\ee
Consistency with the fact that $\ve_-$ does not depend on $r$ requires that

\be
(1 + \G^{2345}) \ve_- =0
\ee
Note that 

\be
\G^+ \ve_- = (1+ \G^{2345}) \ve_- =0 \;\;\; \Rightarrow \;\;\; (1+ \G^{(6)}) \ve_- =0
\ee
so all our equations are consistent. As fas as $\ve_+$, is concerned, if we multiply \eqref{teqn} by $\G^+$ we obtain

\be
(1+ \s^1) \ve = 2 \ve_+ =0
\ee
so we obtain exactly the same Killing spinors as we did for the $SL(2,\mathbb{R})_L \times SU(2)_R$ invariant Schr\"{o}dinger spacetime. This result was expected of course, based on the fact that the corresponding operator deformations \eqref{nso} preserve the same supersymmetries.

\end{document}